\documentclass[11pt]{article}
\usepackage{fix-cm}
\usepackage{amssymb}
\usepackage{amsmath}
\usepackage{graphicx}
\usepackage{subcaption}
\usepackage{makeidx}
\usepackage{multicol}
\usepackage[colorlinks=true,allcolors=blue]{hyperref}
\usepackage[capitalize]{cleveref}
\crefname{figure}{Figure}{Figures}
\usepackage{xcolor}
\usepackage{placeins} 
\usepackage{fancyvrb} 
\usepackage{enumitem}
\usepackage[text={440pt,575pt},headheight=9pt,centering]{geometry}

\usepackage{amsthm}

\newtheorem{definition}{Definition}[section]
\newtheorem{example}{Example}[section]
\newtheorem{remark}{Remark}[section]
\usepackage{mathabx}
\usepackage{authblk}
\usepackage{natbib}

\title{Graphical models for multivariate extremes}
\author[1]{Sebastian Engelke}
\author[2]{Manuel Hentschel}
\author[3]{Micha\"el Lalancette}
\author[4]{Frank R\"ottger}
\affil[1]{
	Universit\'e de Gen\`eve,
	\href{mailto:sebastian.engelke@unige.ch}{\texttt{sebastian.engelke@unige.ch}}
}
\affil[2]{
	Universit\'e de Gen\`eve,
	\href{mailto:manuel.hentschel@unige.ch}{\texttt{manuel.hentschel@unige.ch}}
}
\affil[3]{
	Universit\'e du Qu\'ebec \`a Montr\'eal,
	\href{mailto:lalancette.michael@uqam.ca}{\texttt{lalancette.michael@uqam.ca}}
}
\affil[4]{
	TU Eindhoven,
	\href{mailto:f.rottger@tue.nl}{\texttt{f.rottger@tue.nl}}
}
\date{\today}

\newcommand{\G}{\mathcal{G}} 
\newcommand{\T}{\mathcal{T}} 
\newcommand{\V}{\mathcal{V}} 
\newcommand{\E}{\mathcal{E}} 
\newcommand{\C}{\mathcal{C}} 
\newcommand{\D}{\mathcal{D}} 
\newcommand{\ph}{\text{ph}} 
\renewcommand{\L}{\mathcal{L}} 
\newcommand{\Lm}{\mathcal{L}^{(m)}} 
\DeclareMathOperator{\argmin}{arg\,min} 

\renewcommand{\d}{{d}} 
\newcommand{\bone}{\mathbf{1}} 
\newcommand{\Sym}{\mathcal{S}^\d} 
\newcommand{\Sone}{\Sym_{\bone}} 
\newcommand{\SHR}{\Sym_{\bone, +}} 
\newcommand{\CHR}{\C^\d} 

\newcommand{\y}{\mathbf{y}} 
\newcommand{\Y}{\mathbf{Y}} 
\newcommand{\Ym}{\mathbf{Y}^{(m)}} 

\newcommand{\X}{\mathbf{X}}
\newcommand{\Z}{\mathbf{Z}}
\newcommand{\U}{\mathbf{U}}
\newcommand{\z}{\mathbf{z}}
\newcommand{\PP}{\mathbb{P}}
\newcommand{\RR}{\mathbb{R}}
\newcommand{\EE}{E^+}
\newcommand{\one}{\mathbf{1}}
\newcommand{\zero}{\mathbf{0}}
\newcommand{\dirac}{\delta}
\newcommand{\tr}{{\rm tr}} 
\newcommand{\minimize}{\mathop{\rm minimize}} 
\newcommand{\onenormoff}[1]{\|#1\|_{1, \rm{off}}}

\newcommand{\W}{\mathbf{W}}

\newcommand{\pa}{\text{pa}} 
\newcommand{\neigh}{\text{ne}} 
\newcommand{\des}{\text{des}} 

\newcommand{\set}[1]{\{#1\}}
\newcommand{\setM}[2]{\{#1\mid#2\}}

\newcommand{\EM}{\Lambda} 
\newcommand{\emd}{\lambda} 
\newcommand{\EMF}{V}
\newcommand{\diag}{{\rm diag}}
\newcommand{\eps}{{\varepsilon}}

\newcommand{\MTPtwo}{$\text{MTP}_2$} 
\newcommand{\EMTPtwo}{$\text{EMTP}_2$}

\newcommand{\Prob}{\mathbb{P}} 
\newcommand{\Var}{\text{Var}} 
\newcommand{\Borel}{\mathcal{B}}

\newcommand{\indep}{\perp \!\!\! \perp}
\newcommand{\condInd}[3]{{#1 \indep #2 \mid #3}}
\newcommand{\exCondInd}[4][\EM]{{#2 \perp_{#1} #3 \mid #4}}
\newcommand{\exInd}[3][\EM]{{#2 \perp_{#1} #3}}

\newcommand{\HR}{H\"usler--Reiss}

\newcommand{\eglearn}{EGlearn}

\newcommand{\includePlot}[2][]{
    \includegraphics[#1]{figures/#2.pdf}
}

\newcommand{\IfPlotExists}[3]{\IfFileExists{figures/#1.pdf}{#2}{#3}}

\begin{document}
    
	\maketitle

	\begin{abstract}
  
Graphical models in extremes have emerged as a diverse and quickly expanding research area in extremal dependence modeling.
They allow for parsimonious statistical methodology and are particularly suited for enforcing sparsity in high-dimensional problems.
In this work, we provide the fundamental concepts of extremal graphical models and discuss recent advances in the field. Different existing perspectives on graphical extremes are presented in a unified way through graphical models for exponent measures. We discuss the important cases of nonparametric extremal graphical models on simple graph structures, and the parametric class of \HR{} models on arbitrary undirected graphs. In both cases, we describe model properties, methods for statistical inference on known graph structures, and structure learning algorithms when the graph is unknown.
We illustrate different methods in an application to flight delay data at US airports.

	\end{abstract}

	\section{Introduction}
  \label{ch13:sec:introduction}

Conditional independence for a random vector $\X = (X_i : i \in \V)$ with index set $\V = \{1,\dots,d\}$ forms the basis for many tools of modern statistical inference. A probabilistic graphical model represents the components of the vector $\X$ and their conditional independencies through a graph structure $\G = (\V, \E)$, where variables are indexed by nodes $\V$ and connected by edges $\E$. This notion naturally decomposes the joint distribution into smaller models and thus simplifies  statistical inference.
Both undirected and directed graphical models offer principled and interpretable ways to define sparse statistical models, and they have lead to many successful applications.
For instance, they provide sparse representations of the joint dependence structure of Gaussian models \citep{wainwright2008} and several extensions, such as skew-normal and Gaussian copula models \citep{capitanio2003,LLW09}. Multiple graphical models have been developed for continuous exponential family distributions \citep{inouye2016square,YRAL15}, including dependence models for multivariate angular data \citep{KOBMK20}.
The Ising model, an extensively studied model in particle physics, is a graphical model for multivariate binary data \citep{pra2010}. For more general categorical data, graphical models on log-linear interaction models are commonly used for contingency tables \citep{darroch1980,lauritzen1996}. Other graphical models for discrete data include the discrete square root and Poisson graphical models \citep{inouye2016square,yang2013poisson}.
Bayesian networks encode conditional independence and density factorization via separation statements for directed acyclic graphs \citep{Pearl2000}.
They link to causal inference through structural causal models, where the graphical structure allows for a causal interpretation \citep{maathuis2019handbook, PJS2017}.

Multivariate extreme value theory is concerned with the theoretical understanding and the statistical modeling of the distributional tail of the random vector $\X$.
Mathematically, this tail can be described by three different, equivalent approaches: the point process approach, the block maxima method leading to max-stable distributions, and the threshold exceedances approach with limiting multivariate Pareto distributions.
For all of these methods, for a sample of size $n$, the effective sample size $k\ll n$ is much smaller since only the largest $k$ observations of $\X$ carry relevant information on its tail and are used for inference. Therefore, sparse extreme value models that make efficient use of the data are of utmost importance \citep{engelke2021a}. 
A natural question is thus how powerful tools from graphical modeling can be employed in the framework of statistics of extremes.

Compared to the classical applications of graphical models described above, the focus here is different. Indeed, while probabilistic conditional independence mainly concerns the bulk of the distribution, an appropriate notion of an extremal graphical model should describe the dependence structure in the tail of $\X$. Conceptually, there are at least two different ways of approaching this problem: (a) one can assume that $\X$ is a classical graphical model and study its extremal limit; or (b) one may directly define graphical models for the limiting objects in the three approaches to extremes described above, relying on a tailor-made notion of extremal conditional independence.
Eventually, both of the resulting theories, where they overlap, yield the same graphical models for extremes.

\begin{figure}
	\begin{subfigure}{0.24\textwidth}
		\includePlot[width=1\textwidth]{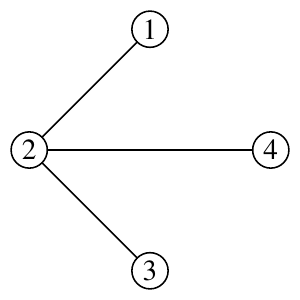}
		\caption{Tree}
		\label{ch13:subfig:graphTree}
	\end{subfigure}
	\begin{subfigure}{0.24\textwidth}
		\includePlot[width=1\textwidth]{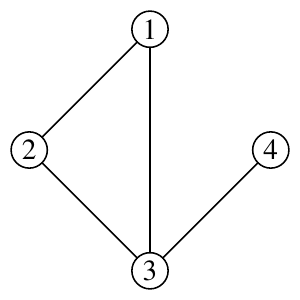}
		\caption{Block}
		\label{ch13:subfig:graphBlock}
	\end{subfigure}
	\begin{subfigure}{0.24\textwidth}
		\includePlot[width=1\textwidth]{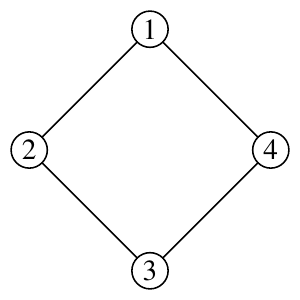}
		\caption{Cycle}
		\label{ch13:subfig:graphCycle}
	\end{subfigure}
	\begin{subfigure}{0.24\textwidth}
		\includePlot[width=1\textwidth]{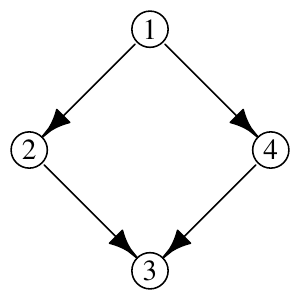}
		\caption{DAG}
		\label{ch13:subfig:graphDAG}
	\end{subfigure}
	\caption{
		Examples of graph structures on the set of nodes $\V = \{1,\dots, 4\}$.
	}
	\label{ch13:fig:intro}
\end{figure}

Perspective (a) goes back to the analysis of extremes of time series, 
where the extremal limit is a multiplicative random walk on a chain graph \citep{smi1992, per1994, bor2003, segers2007multivariate}.
In the same spirit, this random walk theory may be extended to trees \citep{segers2020}, block graphs \citep{asenova2023extremes} and certain directed graphs \citep{segers2022max}; see \cref{ch13:fig:intro} for examples of graphs. A limitation of this perspective is that technical assumptions are required on the distribution of $\X$ to guarantee multivariate regular variation~\eqref{mrv}. Moreover, no results exist that go beyond simple graph structures where the limit would no longer have a random walk structure.

The main difficulty of perspective (b) is that classical conditional independence is not useful for any of the three approaches to extreme value modeling. 
Indeed, considering the point process approach, a random set of points is a non-standard object for graphical modeling. For block maxima, given a max-stable vector $\Z$ with positive continuous density and disjoint subsets $A, B, C \subset \V$, \cite{papastathopoulosStrokorb2016} show that classical conditional independence already implies unconditional independence,
\begin{align}\label{CI_max}
	\condInd{\Z_A}{\Z_B}{\Z_C} \quad \Rightarrow \quad \Z_A \indep \Z_B.
\end{align}
Consequently, no nontrivial
conditional independencies are possible in this model class
and densities of max-stable distributions therefore cannot factorize on a graph into lower dimensional functions.
For threshold exceedances, the multivariate Pareto distribution $\Y$ is defined on a non-product space and classical conditional independence  is not applicable to this random vector since even the support depends on the conditioning event. Consequently, in all three cases, a different notion of  conditional independence is required.
\cite{engelkeHitz2020} introduce the concept of extremal conditional independence for the class of multivariate Pareto distributions, and \cite{engelkeIvanovsStrokorb2022} extend it to a unifying notion involving the so-called exponent measure that covers all of the three approaches.

Since then, a lot of progress has been made at the interface of graphical models and extreme value theory, providing a better understanding of the theory and a tool box of statistical methods. 
In this work we discuss the fundamental ideas of extremal graphical models and give an overview over the statistical methodology. 
While we concentrate on probabilistic graphical models where graphs encode conditional independence properties \citep{lauritzen1996}, there exist also other ways of using graphs to build parsimonious statistical models \citep{lee2018, vet2020,kiriliouk2023xvine}. Our focus is the limiting model perspective (b) under the classical asymptotic dependence framework \citep{res2008}; graphical models for asymptotically independent models are still fairly unexplored \citep{kulik2015,papastathopoulos2017,casey2023decomposable}.

\cref{ch13:sec:graphicalModels} provides background on multivariate extremes and classical graphical models.
We introduce extremal notions of conditional independence and graphical models in \cref{ch13:sec:extremalGraphicalModels} and discuss  implications for the limiting models arising in the three approaches to extremes mentioned above. The remaining sections present fundamental properties and statistical methodology for extremal graphical models, with a focus on threshold exceedances. In particular, we describe estimation methods if the underlying graph structure~$\G$ is known, and algorithms for structure learning if the graph is unknown. \cref{ch13:sec:simpleGraphStructures} focuses on non-parametric methods on simple graph structures, while \cref{ch13:sec:hueslerReiss} allows for general graphs but for the parametric class of \HR{} models, an analogue to Gaussian distributions in extremes. \cref{ch13:sec:application} showcases many of the methods on a flight delay dataset from the \texttt{R} package \texttt{graphicalExtremes} \citep{graphicalExtremes2022}.



	\section{Background}
  \label{ch13:sec:graphicalModels}
  \subsection{Multivariate extreme value theory}
\label{sec:MEVT}

Multivariate extreme value theory studies the dependence between the largest observations of the components of a random vector $\X = (X_i : i\in \V)$ with index set $\V = \{1,\dots, d\}$.
In order to focus only on the dependence structure and abstract away the marginal distribution, we assume throughout that $\X$ is normalized to have standard Pareto margins.
In order to mathematically describe this extremal dependence structure, a common assumption is \index{multivariate regular variation}multivariate regular variation, which requires the existence of a Borel measure $\EM$ on $\EE = [0, \infty]^d \setminus \{\zero\}$, such that for all Borel subsets $A\subset \EE$ bounded away from the origin with $\EM(\partial A) = 0$ the limit
\begin{align}\label{mrv} 
	\lim_{n\to\infty} n\PP(\X / n \in A ) = \EM(A),
\end{align}
exists; see \cite{res2008} for details. We say that $\X$ is in the domain of attraction of 
the \index{exponent measure}exponent measure $\EM$, which then has the homogeneity property $\EM(t B) = t^{-1} \EM(B)$ for any $t>0$ and Borel set $B \subset \EE$. 
If $\EM$ possesses a Lebesgue density $\emd$, then it satisfies the two properties 
\begin{enumerate}[label=(P{{\arabic*}})]
	\item \label{ch13:prop:normalization}
		normalization: for any $m \in \{1,\dots, d\}$ it holds that $\int_{\{y_m > 1\}} \emd(\y) \d \y = 1 $; 
	\item \label{ch13:prop:homogeneity}
		homogeneity: $\emd(t \y) = t^{-d-1} \emd(\y)$ for any $t>0$ and $\y \in \EE$. 
\end{enumerate}
Conversely, any positive function on $\EE$ satisfying these two properties uniquely defines an exponent measure. This can be verified by equating the normalization property \ref{ch13:prop:normalization} to the marginal constraint that is known to characterize the class of all possible spectral measures \citep[Section 8.2.3]{beirlant2006statistics}.
For any subset $I\subset \V$, the marginal exponent measure $\EM_I$ and its density $\emd_I$ are defined by integrating out the components $\V \setminus I$.

While~\eqref{mrv} is a purely mathematical condition, it is in fact equivalent to either of the three main approaches to multivariate extremes, which we describe in the sequel. Let $\X_1,\dots, \X_n$ be independent copies of a random vector $\X$ with standard Pareto margins.
\begin{itemize}
    \item (Point process) As $n\to\infty$, the random point set $\sum_{i=1}^n \dirac_{\X_i / n}$ converges in distribution in the space of point process to the \index{Poisson point process}Poisson point process $\Pi_\EM$ on $\EE$ with intensity measure $\EM$; see \cite[Proposition 3.21]{res2008} for details. 
    \item (Block maximum) As $n\to\infty$, the componentwise maximum $n^{-1} \max_{i=1,\dots,n} \X_i$ converges in distribution to a \index{max-stable distribution}max-stable random vector $\Z$ with distribution function
    \[\PP(\Z \leq \z) = \exp\left\{ - V(\z)\right\}, \quad \z \in  [0, \infty)^\d, \]
	where $V(\y) = \EM(\EE \setminus [\zero, \y])$ for any $\y \in \EE$. 
    \item (Peaks-over-threshold) As the threshold $u\to\infty$, the exceedances $u^{-1} \X \mid \max_{j=1,\dots, \d} X_j > u$ converge in distribution to a \index{multivariate Pareto distribution}multivariate Pareto vector $\Y$ with distribution function 
    \[ \PP(\Y \leq \y) = \frac{\EMF(\min\{\y, \one \}) - \EMF(\y)}{\EMF(\one)}, \quad \y \in \L,\]
	where $\min\{\y, \one\}$ denotes the componentwise minimum.
\end{itemize}

\begin{figure}[h!]
	\centering
	\begin{subfigure}{0.31\textwidth}
		\includePlot[width=\textwidth]{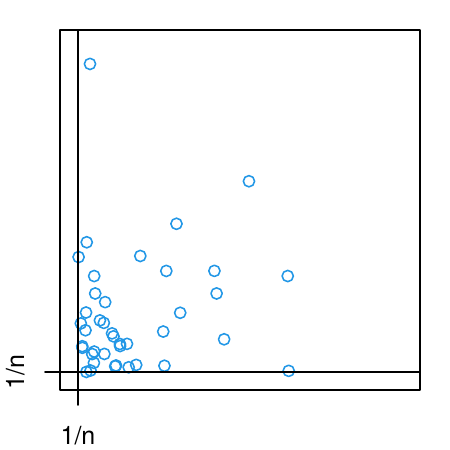}
	\end{subfigure}
	\begin{subfigure}{0.31\textwidth}
		\includePlot[width=\textwidth]{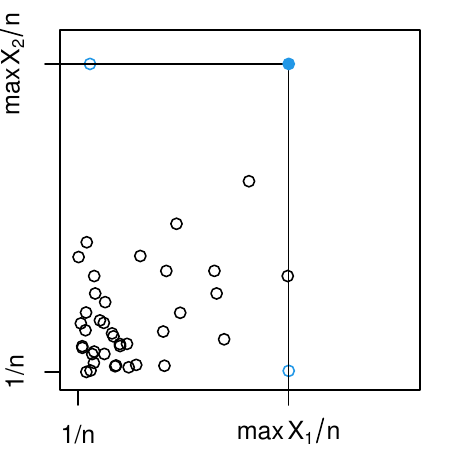}
	\end{subfigure}
	\begin{subfigure}{0.31\textwidth}
		\includePlot[width=\textwidth]{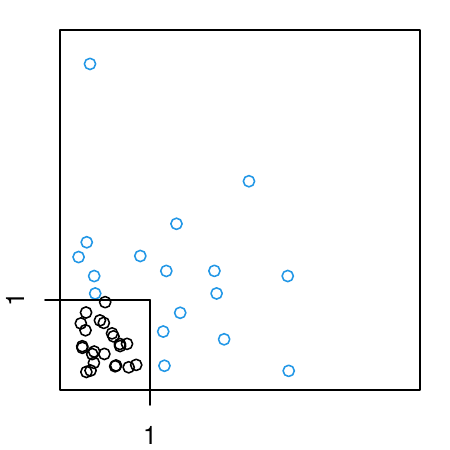}
	\end{subfigure}
	\vspace{-0.5cm}
	\caption{
		All panels show $n$ samples of the random vector $\X$ and illustrate the three main approaches to multivariate extremes:
		the blue points in the left panel are the random point set that is approximated by the Poisson point process $\Pi_\EM$;
		the solid blue point in the center panel is the componentwise maximum with approximate max-stable distribution;
		the blue points in the right panel are the threshold exceedances that are approximate samples of a multivariate Pareto distribution.
	}
	\label{ch13:fig:evtApproaches}
\end{figure}

\begin{remark}
	We emphasize that while the three approaches above appear fairly different, they capture the same information on the tail of the random vector $\X$. Indeed, either of the three limiting objects is in one-to-one correspondence to the exponent measure $\EM$.
	When $\X$ is attracted to $\EM$, we may equivalently say that it is in the domain of attraction of the associated point process $\Pi_{\EM}$, max-stable distribution $\Z$, or multivariate Pareto distribution $\Y$.
	Another equivalent way of representing the extremal dependence in $\EM$ is via the extreme value copula that attracts the copula of $\X$ \citep{gudendorf2010extreme}.
\end{remark}

A widely used summary of extremal dependence is the extremal correlation 
\[ \chi_{ij} = \lim_{u\to \infty} \mathbb P(X_j > u \mid X_i > u) \in [0,1],\]
where the limit always exists if $\X$ is in the domain of attraction of $\EM$.
It can be reexpressed in terms of any of the limiting objects in the above approaches:
\[\chi_{ij} = 2 - V_{ij}(1,1) = 2 - \log \mathbb P(Z_i \leq 1, Z_j \leq 1) = \mathbb P(Y_j > 1 \mid Y_i >1) .\]
The set of $d\times d$ extremal correlation matrices $\chi$ with entries $\chi_{ij}$, $i,j=1,\dots ,d$, is a strict subset of the set of correlation matrices with 
positive entries \citep{fie2017}.

Another aspect of the extremal dependence of $\X$ is given by considering the sub-faces of $\EE$ defined as
\[ \EE_I = \{\y\in \EE \,:\, y_j\neq 0 \,\,\forall\, j\in I\, \text{ and }\,  y_j=0\,\forall\, j\notin I\}, \]
for a subset $I\subset V$. If $\EM(\EE_I) > 0$, then the variables $\X_I$ can be concomitantly extreme while the other variables $\X_{\V\setminus I}$ are small. There are several statistical methods in the literature that aim to detect such sub-faces with mass \citep{chiapino2019identifying,simpson2020determining,meyer2023multivariate}; for a review see \cite{engelke2021a}.

There are numerous parametric and non-parametric models for multivariate extremes.
Any model defined for one of the three approaches above directly induces the associated models for the other two approaches. We discuss some classical models and refer to \cite{gudendorf2010extreme} for a detailed overview.

\index{max-linear model}
\begin{example}[Max-linear model]\label{max_linear}
    Let $\eps_j$, $j =1,\dots, p$ be independent standard Fr\'echet variables and $A=(a_{ij})$
    be a $\d\times p$ matrix of non-negative coefficients. Define a $\d$-dimensional random vector $\Z$ with entries
  \begin{align}  \label{ML_model}
    Z_i=\max_{j=1,\dots, p} a_{ij}\eps_j ,\quad i=1,\dots, \d.
  \end{align}
  Assume that the rows of $A$ sum to~$1$, ensuring that all $Z_i$ have standard Fr\'echet distributions. The max-linear model is max-stable with exponent measure $\EM$ supported on $p$ rays specified by the columns of $A$, that is, the angles of the rays take $p$ possible values $a_{\cdot j}/\|a_{\cdot j}\|$ with mass proportional to $\|a_{\cdot j}\|$. 
\end{example}
The max-linear model is somewhat degenerate since neither the max-stable distribution nor the corresponding exponent measure $\EM$ possess Lebesgue densities. On the other hand, many continuous models that admit densities have been proposed in the literature, such as the extremal logistic \citep{gumbel1960}, the extremal Dirichlet \citep{col1991} or extremal $t$-distribution \citep{schlater2002, opi2013}.

A highly flexible and very popular parametric model is the \index{H\"usler--Reiss distribution}\HR{} model \citep{hueslerReiss1989}. Similarly to a Gaussian distribution, it is parameterized by a symmetric $(d \times d)$-matrix of bivariate dependence parameters. \HR{} distributions also arise as finite-dimensional marginals of Brown--Resnick processes \citep{bro1977, kabluchkoSchlatherDeHaan2009}, which are ubiquitous models in spatial extreme value modeling. This model will be the focus of \cref{ch13:sec:hueslerReiss}.
\begin{example}[H\"usler--Reiss model]
\label{ex:HR}
	The \HR{} model is parameterized by a symmetric, strictly conditionally negative definite matrix $\Gamma$, also called variogram, in the cone
	\[
		\CHR
		=
		\left\{
			\Gamma \in \mathbb [0, \infty)^{d\times d}:		
			\Gamma = \Gamma^\top
			,\;\;
			\diag{\Gamma} = \mathbf{0}
			,\;\;
			v^\top \Gamma v < 0
			\;\forall\,
			\mathbf{0} \neq v \perp \mathbf{1}
		\right\}.
	\]
	It is defined through its exponent measure density
	\begin{equation} \label{eq:HRdensityGamma}
		\emd(\y) \propto \prod_{i=1}^d y_i^{-1-1/d} \exp\Big\{ -\frac12 (\log y - \mu_\Gamma)^\top \Theta (\log y - \mu_\Gamma) \Big\}, \quad \y \in \EE,
  \end{equation}
	where $\Theta = (P(-\Gamma/2)P)^+$, $P = I_d - \bone\bone^\top /d$ is the projection matrix onto the orthogonal complement of $\bone$, $A^+$ is the Moore--Penrose pseudoinverse of a matrix $A$, and $\mu_\Gamma = P (-\Gamma/2) \bone$; see \cite{hentschelEngelkeSegers2022} for details.
\end{example}

\subsection{Conditional independence and graphical models}
\label{sec:CI_background}



Before discussing graph separation and its role in probabilistic graphical models, we summarize the necessary graph theory. We give complete details on undirected graphs, whereas some more technical notions on directed graphs are deferred to Section~\ref{ch13:sec:appendixA}.

A directed \index{directed graph}graph is a pair $\G = (\V,\E)$
consisting of a vertex set $\V$
and an edge set $\E \subseteq \V \times \V$,
containing ordered pairs of distinct vertices; the terms vertex and node are generally used interchangeably.
Without loss of generality we shall consider that $\V = \{1, \dots, d\}$.
An \index{undirected graph}undirected graph is a graph in which the
presence of an edge $(i,j) \in \E$ also implies $(j,i) \in \E$; the two edges are then considered equivalent and graphically represented by a single undirected connection. \cref{ch13:fig:intro} shows examples of undirected graphs in the first three panels and a directed graph in the fourth panel.

A path is a sequence of distinct vertices $(i_0, i_1, \ldots, i_L)$
such that $(i_{k-1}, i_{k}) \in \E$ or $(i_k,i_{k-1}) \in \E$ for all $k = 1, \ldots, L$.
A cycle is a sequence of vertices $(i = i_0, i_1, \ldots, i_L = i)$ such that $(i_{k-1}, i_{k}) \in \E$ or $(i_k,i_{k-1}) \in \E$ for all $k = 1, \ldots, L$, and where no vertex is visited more than once except the start and end point~$i$. 
Paths and cycles are called directed if for each $k$, $(i_{k-1}, i_{k}) \in \E$. 
A directed graph that contains no directed cycles 
is called a \index{directed acyclic graph}directed acyclic graph (DAG).
Two vertices $i, j$ are said to be connected if there exists a (not necessarily directed) path
$(i, \ldots, j)$ between them.
If all pairs of vertices in a graph are connected,
the graph is referred to as connected,
otherwise it can be partitioned into connected components.
If there is an edge $(i,j)$ between vertices $i$ and $j$,
they are called adjacent.
The set of all vertices adjacent to~$i$ is called its neighborhood
and denoted $\neigh(i)$.
A maximal subset of nodes where every pair of nodes is adjacent is called a clique.

In order to connect graphical structures to probability distributions and make them amenable for statistical modeling, the notion of \index{graph separation}separation in a graph is crucial.

\begin{definition}[Undirected separation]\label{def:sep_und}
	Let $\G=(\V,\E)$ be an undirected graph and let  $A,B,C\subseteq \V$ be disjoint sets. The set $A$ is separated from $B$ by $C$ if any path between $A$ and $B$ in $\G$ has a non-empty intersection with $C$.
\end{definition}
We call the smallest set of nodes that separates $A$ from $B$ a minimal separator. An undirected graph where each minimal separator between any pair of connected nodes is a clique is called decomposable. 
For DAGs, the ubiquitous notion of separation is termed \index{$d$-separation}$d$-separation. It is considerably more involved, so its formal definition is deferred to \cref{ch13:sec:appendixA}. In short we write $A\perp_\G B \mid C$ whenever $A$ is graph separated from $B$ by $C$ in an undirected graph or DAG $\G$. When some of these sets are singletons, we typically write $a \perp_\G b \mid c$ for $\{a\} \perp_\G \{b\} \mid \{c\}$.


\begin{example}
	\label{ch13:example:dSeparation}
	
	The graph separation statements implied by the undirected graph in \cref{ch13:subfig:graphCycle} are $1 \perp_\G 3 \mid \{2, 4\}$ and $2 \perp_\G 4 \mid \{1, 3\}$. In contrast, the DAG in \cref{ch13:subfig:graphDAG} does not imply that $2 \perp_\G 4 \mid \{1, 3\}$, but instead that $2 \perp_\G 4 \mid 1$.
\end{example}

Probabilistic graphical models on both undirected graphs and DAGs are stochastic models that are constrained to satisfy certain conditional independence relations. In this section we discuss conditional independence of a random vector $\X = (X_i: i\in \V)$ in the classical sense and write $\X_A\indep\X_B\mid \X_C$ when $\X_A$ is conditionally independent of $\X_B$ with respect to $\X_C$, for subsets $A,B,C \subset\V$; a different notion of conditional independence for extremes is introduced in \cref{ch13:sec:extremalGraphicalModels}.
The conditional independencies are determined by through a Markov property that links them to graph separation statements in $\G$.

\index{global Markov property}
\index{probabilistic graphical model}
\begin{definition}[Global Markov property and graphical models]
We say that a random vector $\X$ indexed by $\V$ satisfies the global Markov property with respect to the undirected graph or DAG $\G$ when
\[A\perp_\G B\mid C \quad \Rightarrow \quad \X_A\indep\X_B\mid \X_C\]
for all disjoint $A,B,C\subseteq \V$. In this case, we call $\X$ a (probabilistic) graphical model with respect to graph $\G$.
\end{definition}

While the global Markov property allows a unified definition of graphical models, it is customary and often more convenient to utilize other Markov properties that are equivalent under mild assumptions.
For example, for an undirected graph $\G=(\V,\E)$ the \index{pairwise Markov property}pairwise Markov property requires that
\[
	X_i\indep X_j\mid \X_{\V\setminus\{i,j\}},\quad \forall (i,j)\not\in\E.
\]
It is equivalent to the global Markov property on $\G$ under the assumption of a positive density with respect to a product measure, for instance.
We refer to \cite{lauritzen1996} for a discussion of this and other Markov properties.

Both conditional independence and graph separation are ternary relations for disjoint index sets $A,B,C\subseteq \V$.
Such ternary relations have been generalized to a more abstract notions of independence, which can be studied through axiomatization.
It has been shown that conditional independence of probability distributions, undirected graph separation and $d$-separation for DAGs all conform with a popular collection of axioms called a semi-graphoid \citep{pearl1988,lauritzen1996,GVP1990}.
For more details we further refer to \cite{lauritzen1996,LS2018}.

	\section{Graphical modeling in extremes}
	\label{ch13:sec:extremalGraphicalModels}

\subsection{Extremal graphical models}

\index{Extremal conditional independence}Extremal conditional independence was originally defined in \cite{engelkeHitz2020} as an extension of classical conditional independence to the framework of threshold exceedances. It was used to define extremal graphical models for multivariate Pareto distributions that admit a density. 
In \cref{sec:MEVT} we discussed that the extreme observations of some random vector $\X$ can be described by the corresponding exponent measure $\EM$, regardless of whether the perspective of point processes, block maxima or threshold exceedances is taken. 

We therefore follow the more general notion of extremal conditional independence on the level of the exponent measure $\EM$ in \cite{engelkeIvanovsStrokorb2022}. This approach does not require the existence of densities and naturally encompasses all three perspectives on extremes. The challenge is that $\EM$ is not a probability measure since it explodes at the origin and therefore has infinite mass on the space $\EE$. To circumvent this issue, we define the set of all rectangles with positive $\EM$-mass that are bounded away from the origin as 
\begin{align*}
    \mathcal R(\EM)=\bigg\{R = [\mathbf a, \mathbf b] \, : \mathbf a \leq \mathbf b, \,  \EM(R)>0, \, \zero \notin  R\bigg\}.
\end{align*}
The following definition of conditional independence for $\EM$ essentially requires the corresponding independence statement to hold for any restriction of $\EM$ to rectangles with finite $\EM$-mass.


\begin{definition}
    \label{ch13:def:extremalCondInd}
    Let $A,B,C \subseteq \V$ be a partition of $\V$.
    The exponent measure $\EM$ is said to admit extremal conditional independence of $A$ and $B$ given $C$, 
    \begin{align}\label{extCI}
         \exCondInd{A}{B}{C}, 
    \end{align}
    if we have classical conditional independence on any restriction of $\EM$ on rectangles bounded away from the origin, that is,
    \begin{align}\label{test_class}
         \condInd{\U_A}{\U_B}{\U_C} \text{ for } \U \sim  \EM(\cdot) / \EM(R) \quad \text{ for all } R\in\mathcal R(\EM).
    \end{align}
    This is trivially true for $A$ or $B$ being empty, and for $C=\emptyset$ we say that $\EM$ admits extremal independence of $A$ and $B$, and write \[\exInd{A}{B}.\]
    If the sets $A$, $B$ and $C$ are not a partition of $\V$, then the above definition remains the same with the test class in~\eqref{test_class} replaced by $ \mathcal R(\EM_{A\cup B\cup C})$.
\end{definition}

This definition of conditional independence is tailor-made for extremes and turns out to be very natural. First, it satisfies the so-called semi-graphoid axioms \citep{lauritzen1996}, a set of desirable criteria for conditional independence notions; see \cite{engelkeIvanovsStrokorb2022} for details. 
Second, if $\EM$ admits a positive Lebesgue density $\emd$, then~\eqref{extCI}  is equivalent to the factorization of this density into $\emd(\y) = \emd_{A\cup B}(\y_{A\cup B})\emd_{B\cup C}(\y_{B\cup C}) / \emd_C(\y_C)$ for all $\y \in \EE$.

It is worthwhile to note that extremal conditional independence has different properties than those known from the classical, probabilistic notion for random vectors.
For instance, the case of extremal independence $\exInd{A}{B}$ between two sets $A$ and $B$ that form a partition of $\V$ is not equivalent to the factorization of the exponent measure density into the marginal densities; in fact, in this case, a Lebesgue density of $\EM$ can not exist. Instead, extremal independence is equivalent to the fact that $\EM$ has only mass on the sub-faces $\EE_A$ and $\EE_B$, that is,
\begin{align}\label{extInd}
     \exInd{A}{B} \quad \Leftrightarrow \quad \EM(\y_A \neq \zero_A \text{ and } \y_B \neq \zero_B) = 0. 
\end{align}
This fact is highly desirable since it means that this notion is compatible with the traditional notion of asymptotic independence; indeed, the latter is said to hold for sub-vectors $\X_A$ and $\X_B$ of a random vector $\X$ in the domain of attraction of $\EM$ whenever the right-hand side of~\eqref{extInd} is satisfied \citep{strokorb2020extremal,engelkeIvanovsStrokorb2022}.
As a consequence of~\eqref{extInd}, the existence of a density $\emd$ implies the graph $\G$ to be connected, a surprising property of extremal graphical models.

We now define \index{extremal graphical model}extremal graphical models through an extremal global Markov property, which links graph separation for undirected graphs and DAGs (Definitions~\ref{def:sep_und} and~\ref{def:sep_dir}) with the notion of extremal conditional independence. 
\begin{definition}\label{def:global_Markov}
    The exponent measure $\EM$ satisfies the global Markov property with respect to an undirected graph or DAG $\G = (\V, \E)$ if 
    \begin{align*}
    	A\perp_\G B \mid C \quad \Rightarrow \quad \exCondInd{A}{B}{C}
    \end{align*}
    for any disjoint  $A, B, C\subset \V,$ where a (possibly empty) set
    $C$ separates $A$ from $B$. In this case, we say that $\EM$ is an extremal graphical model with respect to~$\G$. 
\end{definition}

This definition includes both undirected graphical models and directed graphical models. For those two classes, we next discuss some fundamental properties and give examples. 

\subsubsection{Undirected extremal graphical models}

If $\G$ is an undirected graph, the global Markov property implies the pairwise Markov property
\[ (i,j) \notin \E \quad \Rightarrow \quad \exCondInd{i}{j}{V\setminus\{i,j\}}, \] 
which allows us to read off extremal conditional independencies directly from missing edges in $\G$. 
Extremal graphical models also contain information on which variables can be extreme at the same time. More precisely, for any $D\subset \V$ such that the sub-graph of $\G$ restricted to $D$ is disconnected, it holds that $\EM{}(\EE_D) = 0$ \cite[Corollary 6.3]{engelkeIvanovsStrokorb2022}.

If the exponent measure $\EM$ admits a positive Lebesgue density on $\EE$ then pairwise and global Markov properties are equivalent. If the graph $\G$ is decomposable then, by a Hammersley--Clifford theorem, this is further equivalent to the factorization of the exponent measure density into marginal densities on the cliques $\C$ of the graph 
\begin{align}
    \label{HC}
    \emd(\y) = \prod_{C\in \C} \emd_C(\y_C) \Big/ \prod_{D\in \D} \emd_D(\y_D), \quad \y \in \EE,
\end{align}
where $\D$ is a multiset containing intersections between the cliques called separator sets \cite[Theorem 1]{engelkeHitz2020}. This property is the basis for efficient statistical modeling and inference, by specifying and estimating the lower-dimensional exponent measure densities associated to cliques and separator sets \citep{engelkeHitz2020}.


\begin{example}
	Let $\EM$ be an undirected extremal graphical model with respect to the decomposable graph in Figure~\ref{ch13:subfig:graphBlock} that admits a positive Lebesgue density $\emd$. Then for all $\y \in \EE$,
	\[\emd(\y)=\emd_{123}(\y_{123}) \emd_{34}(\y_{34}) / \emd_{3}(y_{3}).\]
\end{example}



\subsubsection{Directed extremal graphical models}

If $\EM$ is an extremal graphical model with respect to a DAG $\G$, the global Markov property again implies an extremal directed factorization property.
In particular, if an extremal graphical model $\EM$ admits a positive Lebesgue density $\emd$, then this density factorizes on $\G$ as 
\begin{align}\label{density_factorization}
    \emd(\y) & = \prod_{v\in \V} \emd(y_v\mid \y_{\pa(v)}), \quad \y \in \EE,
\end{align}
where we define the conditional exponent measure density as $\emd(y_v\mid \y_{\pa(v)}) = \emd(y_v,\y_{\pa(v)}) /\emd(\y_{\pa(v)})$ and $\pa(v)$ denotes the parent set of node $v$; see \cref{ch13:sec:appendixA}.
We note that by the normalization property \ref{ch13:prop:normalization} in \cref{sec:MEVT} the conditional density $\emd(\cdot \mid\y_{\pa(v)})$ is a probability density for almost all $\y_{\pa(v)} \in [0,\infty)^{|\pa(v)|}$ and that, by the homogeneity property \ref{ch13:prop:homogeneity} it satisfies
\begin{align}\label{cond_hom}
    \emd(t y_v \mid t \y_{\pa(v)}) = t^{-1}\emd(y_v\mid \y_{\pa(v)}),
\end{align}
for any $\y \in \EE$ and $t > 0$.

\begin{example}
	Let $\EM$ be an extremal graphical model with respect to the DAG in Figure~\ref{ch13:subfig:graphDAG} that admits a positive Lebesgue density $\emd$.
	Then,
	\[\emd(\y)=\emd(y_1)\emd(y_2|y_1)\emd(y_4|y_1)\emd(y_3|y_2,y_4), \quad  \y \in \EE.\]
\end{example}

Extremal graphical models imply factorizations of the exponent measure density $\emd$ as in~\eqref{HC} and~\eqref{cond_hom}. This directly links to the point process approach to extremes since the density $\emd$ is the intensity of the limiting Poisson point process $\Pi_{\EM}$ in this framework. In the next two sections we discuss the implications of extremal conditional independence and graphical models for the two other approaches: the peak-over-threshold and the block maxima method.

\subsection{Multivariate Pareto distributions}

Of the three approaches to extreme values described in Section~\ref{sec:MEVT},
the framework of threshold exceedances and their limiting multivariate Pareto distributions $\Y$ is the most naturally suited to graphical modeling.
This is due to the fact that their probability measure is proportional
to the corresponding exponent measure, in the sense that
\begin{align*}
    \Prob(\Y \in A) = \EM(A) / \EMF(\bone)
    ,
    \quad 
    A \in \Borel(\L),
\end{align*}
and, if it exists, so is the probability density function
$f(\y) = \emd(\y) / \EMF(\bone)$.
In view of~\eqref{HC}, 
if $\EM$ is an extremal graphical model on the undirected graph $\G$,
the density $f$ factorizes on $\G$.

Since the support of $\Y$ is not a product space, we need to restrict 
the random vector to suitable rectangles to obtain classical conditional independence statements. For $m\in \V$, choose the rectangle $\Lm = \setM{x \in \L}{x_m > 1}$ in \cref{ch13:def:extremalCondInd} and note that the corresponding random vector with law $\EM(\cdot) / \EM(\Lm)$ has the same
distribution as 
\begin{align}\label{Ym}
    \Ym = \Y \mid \set{\Y_m > 1}.
\end{align}
This random vector has the interpretation as the limit when the $m$th component is extreme, and it exhibits classical conditional independence according to the underlying graph $\G$; in fact, this was the original definition of extremal conditional independence in~\cite{engelkeHitz2020}.
Working with $\Ym$ often allows us to apply
statistical methods tailored to classical conditional independence models.
Together their supports cover the entire set $\L$ and a joint estimator based on observations from each $\Ym$, $m\in \V$, uses all observations from $\Y$ itself. An important summary statistic based on these random vectors is the extremal \index{extremal variogram}variogram rooted at node $m\in V$ defined as the matrix $\Gamma^{(m)}$ with entries
\begin{align}\label{ext_vario}
     \Gamma^{(m)}_{ij} = \Var\left\{\log Y_i^{(m)} - \log Y_j^{(m)} \right\}, \quad i,j \in V,
\end{align}
whenever the right-hand side is finite \cite[Section 3]{engelkeVolgushev2022}. For many parametric models, this matrix has a one-to-one correspondence to the model parameters, as for instance for the logistic and the \HR{} models.

\begin{example}\label{Ym_HR}
    If $\Y$ follows a \HR{} multivariate Pareto distribution with parameter matrix $\Gamma \in \mathcal C^d$, for $m\in \V$, then the random vector in~\eqref{Ym} has the representation 
    \begin{align}\label{Y_normal}
        \left(\log Y^{(m)}_{i} -  \log Y^{(m)}_m\right)_{i\neq m}  \sim N(-  \text{diag}(\Sigma^{(m)})/2, \Sigma^{(m)}),
    \end{align}
    where $\Sigma^{(m)}$ is a $(d-1)\times (d-1)$-dimensional positive definite covariance matrix obtained from $\Gamma$ via $\Sigma_{ij}^{(m)}=\frac{1}{2}\left(\Gamma_{im}+\Gamma_{jm}-\Gamma_{ij}\right)$ for $i, j \neq m$. This implies that for \HR{} distributions, all extremal variograms rooted at nodes $m\in \V$ coincide and are equal to the parameter matrix, that is, 
    $$\Gamma = \Gamma^{(1)} = \dots = \Gamma^{(d)}.$$
\end{example}

The extremal conditional independence $\exCondInd{\Y_A}{\Y_B}{\Y_C}$ has another intuitive interpretation in terms of classical conditional independence \citep{hentschelEngelkeSegers2022}. Suppose that we observe an extreme event of any variable in $C$, that is, $\max_{j\in C} Y_j > 1$. Then, conditionally on $\Y_C$, the sub-vectors $\Y_A$ and $\Y_B$ are independent in the usual sense. 
For an extremal graphical structure $\G$, this translates into an extremal local prediction property of the form
\[
	\left(Y_i \mid \Y_{\setminus \{i\}} = \y_{\setminus \{i\} }\right)	
	\stackrel{d}{=}
    \left(Y_i \mid \Y_{\neigh(i)} = \y_{\neigh(i)}\right),
	\qquad
	\y \in \EE
    , \;
    \max_{j \in \neigh(i)} y_j > 1
    ,
\]
where $\neigh(i)$ denotes the neighborhood of vertex $i$ in $\G$; see \cref{ch13:sec:graphicalModels}. Thus, knowing that an extreme event has occurred in the neighborhood of $Y_i$, it suffices to know the neighbors of $Y_i$ for predicting its value.

\subsection{Max-stable distributions}

Following~\eqref{CI_max} in the introduction, max-stable distributions seem not to be suited for graphical modeling since their densities can only factorize trivially.
On the other hand, suppose that the exponent measure satisfies the extremal conditional independence $\exCondInd{A}{B}{C}$. What does this mean for the corresponding max-stable distribution? While there is no density factorization, this statement still introduces sparsity in the max-stable distribution function that is characterized by the exponent measure $\EM$. This allows the natural construction of parsimonious max-stable statistical models and therefore simplifies statistical inference. Moreover, in virtually all exact simulation methods for a max-stable vector $\Z$, many samples from densities proportional to the exponent measure density $\emd$ have to be drawn \citep{die2015,dombry2016,zhi2019}. Sparsity in form of extremal conditional independence statements of $\EM$ therefore speed up simulation significantly \citep[Section 5.4]{engelkeHitz2020}. 


Importantly, implication~\eqref{CI_max} only holds for continuous distributions $\Z$, and consequently, a max-stable distribution that does not admit a density may allow for nontrivial conditional independence structures.
The authors in \cite{gissibl2018} therefore define \index{recursive max-linear model}recursive max-linear models on a DAG $\G$ by
\begin{equation} \label{ML_model_rec}
	Z_i = \max_{k \in \pa(i)} \max\{c_{ki} Z_k, c_{ii} \eps_i\}, \quad i = 1, \dots, d,
\end{equation} 
with independent, positive noise terms $\eps_i$ and edge weights $c_{ki}>0$, $i \in V$, $k \in \pa(i) \cup \{i\}$. The maximum over an empty set is defined as $0$ and thus, for a source node $i$ for which $\pa(i)$ is empty, we have $Z_i = c_{ii} \eps_i$.
While \cite{gissibl2018} do not fix the noise distribution, we follow later papers and assume that $\eps_i$ are standard Fr\'echet distributed. The random vector $\Z$ in \eqref{ML_model_rec} is then a special case of the max-linear model in \cref{max_linear} with $p=d$ noise terms and the coefficient matrix $A$ is defined by $a_{ii} = c_{ii}$ and
\[
    a_{ij} = c_{jj} \max_{ \omega \in P_{ji}}  \prod_{(u,v) \in \omega} c_{uv}
    \quad i,j\in V, i \neq j,
\] 
\citep[Theorem 2.2]{gissibl2018}, where $P_{ji}$ is the set of all directed paths from $j$ to $i$ in the DAG $\G$; here a path $\omega$ is to be understood as a set of edges and the product ranges over these edges.
While the coefficient matrix $A$ is fully identifiable from the joint distribution of $\X$, the DAG and edge weights in the formulation of \eqref{ML_model_rec} are not: different graph structures and sets of edge weights can lead to the same coefficient matrix $A$, hence the same observational distribution \citep{gissibl2021identifiability}.
A notable exception is the case where $\G$ is a directed tree. Its structure is then fully identifiable and can be learned via Chu--Liu/Edmonds' algorithm \citep{tran2021estimating}.

Nevertheless, for recursive max-linear models on an arbitrary DAG, \cite{klu2021scaling} describe algorithms to determine a causal order implied by the unknown graph~$\G$, and to learn this order by estimating the max-linear coefficients $a_{ij}$ using scalings. Building on this, \cite{krali2023heavy} determine sufficient conditions for the edge weights $c_{ki}$ to be identifiable even in the presence of unobserved variables in the graph, a problem reminiscent of that considered in \cite{gneccoEtAl2021} for heavy-tailed linear models.

\begin{figure}
    \centering
    \includePlot[width=0.5\textwidth]{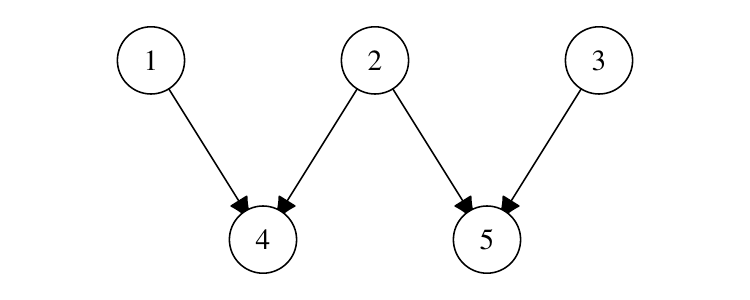}
    \caption{Cassiopeia graph}
    \label{ch13:fig:cassiopeia}
\end{figure}

Like all structural equation models on DAGs the model \eqref{ML_model_rec} satisfies the global Markov property with respect to $\G$, so $\condInd{\Z_A}{\Z_B}{\Z_C}$ whenever $A$ and $B$ are $d$-separated by $C$ in $\G$ \citep[Theorem~1.4.1]{Pearl2000}. However, due to the special structure of recursive max-linear models, they satisfy additional conditional independence relations that are not enforced by $d$-separation. In fact, \cite{amendola2022conditional} show that conditional independence in such models is characterized by $*$-separation, a graphical criterion strictly weaker than $d$-separation.
This is illustrated by the Cassiopeia graph~$\G$ in \cref{ch13:fig:cassiopeia}; while nodes $1$ and $3$ are $d$-connected relative to $\{4, 5\}$, they are $*$-separated by $\{4, 5\}$, and any recursive max-linear model on $\G$ satisfies $\condInd{Z_1}{Z_3}{\Z_{\{4,5\}}}$. 
Interestingly, $*$-separation and $d$-separation share the same Markov equivalence classes \citep{amendola2021markov}, meaning that two DAGs imply the same conditional independence relations on arbitrary distributions if and only if they do so in recursive max-linear models. 

Since recursive max-linear models are max-stable distributions, it
is interesting to understand the relationship between extremal conditional independence statements on the corresponding exponent measure $\EM$ in the sense of \cref{ch13:def:extremalCondInd}, and conditional independence in the classical sense discussed above.
As a partial result in this direction, \cite{engelkeIvanovsStrokorb2022} show the exponent measure of any recursive max-linear model is also a directed extremal graphical model on the same DAG in the $\EM$ sense. On the other hand, it is still unknown whether $d$-separation fully characterizes extremal conditional independence or if it is strictly stronger. It is also an open question whether in these models, extremal conditional independence is stronger than, weaker than, equivalent to (or none of the above) $*$-separation.

	\section{Simple graph structures}
	\label{ch13:sec:simpleGraphStructures}
	\subsection{Properties}

The complexity of the graph $\G = (\V, \E)$ determines the difficulty of statistical inference for the corresponding extremal graphical models. Here, complexity does not only refer to the number of edges $|\E|$, but also to the structural properties of the graph $\G$. 
In this section we discuss simple graph structures, which often allow statistical methods without any parametric assumption on the exponent measure $\EM$.

\begin{figure}
	\centering
	\begin{subfigure}{0.3\textwidth}
		\vspace{0.5cm}
		\includePlot[width=1\textwidth]{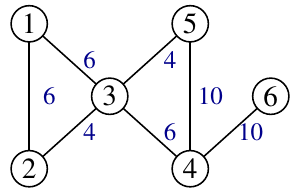}
	\end{subfigure}
	\hspace{0cm}
	\begin{subfigure}{0.6\textwidth}
		\begin{align*}
			\Gamma
			=
			\left(\input{figures/simpleBlockGamma.tex}\right)
		\end{align*}
	\end{subfigure}
	\caption{
		Example of a block graph and an associated extremal variogram
		of a \HR{} distribution.
		Edges are labelled with their corresponding variogram entry.
		Underlined entries, corresponding to non-edges,
		can be computed using \eqref{tree_metric},
		for example
		$\Gamma_{16} = 6 + 6 + 10 = 22$.
	}
	\label{ch13:fig:GammaCompletionBlock}
\end{figure}

A \index{tree}tree $\T = (\V, \E)$ with nodes $\V$ and edge set $\E$ is a connected undirected graph without cycles, such as in \cref{ch13:subfig:graphTree}.
We say that $\EM$ is an extremal tree model if it is an extremal graphical model with respect to a tree.
Trees are the sparsest connected graphs; they contain $|\E| = d-1$ edges, which is much smaller than the number $d(d-1)/2$ of all possible edges in an undirected graph.
A convenient property of trees is the fact that between any two nodes $h,l \in \V$ there is a unique path on $\T$ denoted by $\ph(hl; \T)$,
again to be understood as a set of edges.
To get a first intuition on how the graph structure implies certain properties of the extremal dependence structure, we can consider the extremal correlation coefficient. For an extremal tree model $\EM$ on a tree $\T$ it can be shown \cite[Proposition 5]{engelkeVolgushev2022} that 
\begin{align}\label{chi_tree}
	\chi_{hl} \leq \chi_{ij} \quad \forall (i,j) \in \ph(hl; \T).
\end{align}
This result means that extremal dependence decays with the distance on the tree: the closer two nodes on a path on the tree, the larger the extremal correlation and the stronger the dependence.
These inequalities between different extremal correlations can be used to learn the tree structure in a data driven way. It turns out however, that other coefficients, namely the extremal variograms $\Gamma^{(m)}$ as defined in \eqref{ext_vario}, are much more natural for this purpose. The reason is that they contain information on the underlying tree more directly through the tree metric property 
\begin{align}\label{tree_metric}
	\Gamma_{hl}^{(m)} = \sum_{(i,j) \in \ph(hl; \T)} \Gamma_{ij}^{(m)},
\end{align}
that is, the entry $\Gamma_{hl}^{(m)}$ of two non-adjacent nodes can be computed by summing up the entries of $\Gamma^{(m)}$ on the path between these nodes \citep{engelkeHitz2020,asenova2021inference,engelkeVolgushev2022}.
Consequently, the whole matrix $\Gamma^{(m)}$ is determined by the $d-1$ entries corresponding to the edges $\E$ of the tree; this principle is illustrated in \cref{ch13:fig:GammaCompletionBlock}. As long as the extremal variograms exist, this property is independent of any parametric assumptions on the exponent measure.

The above results suggest that for trees, the bivariate properties completely imply the multivariate dependence structure. This is formalized in terms of the density of the exponent measure which factorizes by~\eqref{HC} into the bivariate marginal densities \cite[Theorem 1]{engelkeHitz2020} as
\begin{align}\label{tree_fact}
	\emd(\y) = \prod_{(ij)\in \E} \frac{\emd_{ij}(y_i, y_j)}{y_i^{-2}y_j^{-2}} \prod_{i\in V} y_i^{-2}\quad \y\in \EE,
\end{align}
where the terms $y_i^{-2}$, $i\in \V$, correspond to the univariate marginal distributions of $\emd$.
A valid extremal tree model on $\T$ can therefore be constructed by choosing, for each edge $(i,j) \in \E$, an arbitrary bivariate exponent measure density (e.g., logistic, \HR{}) and combining them as described above. 
For chain graphs, a subset of tree structures corresponding to Markov chains, this modeling approach was proposed by \cite{col1991,smi1997}.

A slightly more general class than trees are \index{block graph}block graphs, which 
are defined as connected, decomposable graphs where the separator sets are singletons; see \cref{ch13:subfig:graphBlock} for an example. In particular, there is a unique shortest path
between any two nodes of a block graph.
Therefore, some of the properties above generalize naturally to this class of graphs, using this unique shortest path in place of the unique path.
For instance, a factorization similar to~\eqref{tree_fact} allows the construction of parametric extremal block graph models by specifying exponent measure densities on the blocks \cite[Section 5.1]{engelkeHitz2020}. 
Moreover, the additivity property of extremal variograms has been shown to continue to hold if each sub-model on the blocks is \HR{} \citep{engelkeHitz2020,asenova2023extremes}.

Trees and block graphs can be seen as simple ways of combining lower-dimensional distributions into a sparse, higher-dimensional model. For this reason, they are a natural starting point to go beyond multivariate regular variation and build sparse asymptotic independence models. 
For instance, extremes of graphical models that allow for asymptotic independence have been studied in \cite{papastathopoulos2017,papastathopoulos2023hidden} for Markov chains and for trees and block graphs in \cite{casey2023decomposable}. \cite{engelkeIvanovsStrokorb2022} use the general $\EM$ conditional independence in \cref{ch13:def:extremalCondInd} to define a \HR{} tree model which allows for mass on sub-faces of $\EE$, and they propose a construction principle for asymptotic independence trees.

\subsection{Parameter estimation}
\label{ch13:sec_para_est_simple}

Parameter estimation for trees and block graphs is particularly easy. Instead of estimating a full $|\V|$-dimensional model, we can estimate the lower-dimensional models on the cliques of the graph and then combine them through the Hammersley--Clifford theorem; recall that the cliques in trees are all two-dimensional.
Parameter estimation for extremal block graph models therefore boils down to the estimation of low-dimensional, parametric extreme value models. The latter is a classical problem in extreme value statistics and many statistical methods have been proposed.

We concentrate here on the non-parametric, empirical estimator of the extremal variogram $\Gamma^{(m)}$ in~\eqref{ext_vario}, since it will serve as input for most graph structure learning algorithms.
Let $\X_1,\dots,\X_n$ be independent copies of the $d$-dimensional random vector $\X$ in the domain of attraction of a multivariate Pareto distribution $\Y$. Based on the idea of extremal increments \citep{heffernan2004, eng2014, engelke2014}, the \index{empirical variogram}empirical variogram $\hat\Gamma^{(m)}$ is a method of moments estimator with entries 
\begin{equation}\label{emp_vario}
	\hat\Gamma_{ij}^{(m)} =
	\widehat{\Var}\Big(\log (1 - \hat F_i(X_{ti})) - \log(1 - \hat F_j(X_{tj})) : \hat F_m(X_{tm}) \geq 1 - k/n  \Big),
\end{equation}
where $\widehat{\Var}$ denotes the sample variance, $k = k_n$ is an intermediate sequence and $\hat F_i$ is the empirical distribution function of $X_{1i},\dots,X_{ni}$. A close look at this estimator reveals that, first, all marginals are empirically normalized to standard Pareto distributions through the transformations $1/(1-\hat F_i)$. Then, $k$ exceedances in the $m$th component are selected to obtain approximate samples of $\Y^{(m)}$, and finally these samples are used to empirically estimate the variance in~\eqref{ext_vario}.
Under the assumption $k/n \to 0$ as $n\to\infty$ and mild conditions on the underlying data generation, this estimator is consistent for $\Gamma_{ij}^{(m)}$ \citep{engelkeVolgushev2022}, and enjoys subexponential concentration properties \citep{engelke2022a}.
In many applications it makes sense to consider the joint estimator 
\begin{align}
	\label{emp_vario_joint} \hat \Gamma  = \frac1d \sum_{m=1}^d \hat \Gamma^{(m)},
\end{align}
in order to combine information from exceedances in all variables.

For block graphs, for each clique $C\subset \V$, the empirical extremal variogram $\hat \Gamma_{CC}$ of the sub-model can be estimated using only the observations of the components $\X_C = (X_i:i\in C)$, and in many parametric models this directly implies estimates of the parameters. Alternatively, \cite{engelkeHitz2020} propose \index{maximum likelihood estimation}maximum likelihood estimation for the parameters on each clique, and \cite{asenova2021inference} use the pairwise extremal coefficients estimator of \cite{ein2018}. We stress again that any method from extreme value statistics can be used here, the graphical structure providing us with a way to combine the estimated models on the cliques to a valid, $d$-dimensional model on the block graph.

\subsection{Structure learning}
\label{ch13:sec_structure_simple}

The particular properties of trees allow for more efficient statistical structure learning methods than in the case of general graphs. 
Most of these methods rely on the notion of the \index{minimum spanning tree}minimum spanning tree. For a set of symmetric weights $\rho_{ij} > 0$ associated with each pair of nodes $i,j\in \V$, $i\neq j$, the latter is defined as the tree structure
that minimizes the sum of distances on that tree, that is,
\begin{align}\label{Tmin}
	\T_{\text{mst}} = \mathop{\argmin}_{\T = (\V,\E)} \sum_{(i,j)\in \E} \rho_{ij},
\end{align}
where $\T$ ranges over all spanning trees on $\V$.
Given the set of weights, there exist the well-known greedy algorithms by Prim \citep{pri1957} and Kruskal \citep{kruskal1956shortest}  that constructively and efficiently solve this minimization problem.
The crucial ingredient for minimum spanning trees are the weights $\rho_{ij}$, which ideally should be chosen in such a way that $\T_{\text{mst}}$ recovers the true underlying tree structure that represents the conditional independence relations.
In the classical case of Gaussian graphical models, such tree recovery is possible by using weights $\rho_{ij}  = \log (1-r_{ij}^2)/2$, where $r_{ij}$ are the correlation coefficients \citep{drt2017}.
Interestingly, the assumption of Gaussianity is crucial and the result no longer holds outside this specific parametric class. 

For extremal tree models, consistent tree recovery turns out to hold more generally without the need for any parametric assumption on the model class.
Indeed, there are two natural summary statistics for the strength of extremal dependence as candidates for weights in the minimum spanning tree: the extremal correlation in~\eqref{chi_tree} and the extremal variogram in~\eqref{tree_metric}. 
Suppose that $\Y$ is an extremal tree model on the tree $\T$ and let $\hat \chi$ and $\hat \Gamma$ be the empirical estimators based on $n$ observations of $\X$ in the domain of attraction of $\Y$. We further denote by $\hat \T_\chi$ and $\hat \T_\Gamma$ the minimum spanning trees with weights $\rho_{ij} = -\log \hat \chi_{ij}$ and $\rho_{ij} = \hat \Gamma_{ij}$, respectively.
Then, under the conditions for consistency of the empirical estimators, we have
\begin{align}\label{tree_recov}
	\mathbb P( \hat \T_\chi = \T) \to 1, \quad \mathbb P( \hat \T_\Gamma = \T) \to 1, \quad n\to\infty;
\end{align}
for details see \cite[Theorem 2]{engelkeVolgushev2022}.
This result is surprising since it guarantees non-parametric tree recovery, which is stronger than in the non-extreme setting. Let us give some intuition why this is possible. A tree graphical model is characterized by the bivariate distributions on the edges $(i,j) \in \E$. In the case of extremal tree models, these bivariate distributions are represented by the bivariate exponent measure densities $\emd_{ij}(y_i,y_j)$ in the factorization~\eqref{tree_fact}. The crucial property of these densities is the homogeneity described in \cref{sec:MEVT}, which essentially reduces this bivariate distribution to the univariate distribution of the angular part. This implies that any extremal tree model has a random walk structure on the underlying tree $\T$ as shown in \cite[Theorem 1]{segers2020} and \cite[Proposition 1]{engelkeVolgushev2022}. This explains why tree learning is easier in the extremal than in the non-extreme setting. 

The tree recovery result in~\eqref{tree_recov} can be further extended to the setting where the number $d(n)$ of nodes in the tree grows with the sample size.  
In particular, based on concentration bounds for the empirical variogram $\hat\Gamma$ in \cite{engelke2022a}, \cite[Theorem 4]{engelkeVolgushev2022} show that extremal trees can still be consistently learned in high dimensions where $d(n)$ grows exponentially faster than $k$.

	\section{H\"usler--Reiss graphical models}
	\label{ch13:sec:hueslerReiss}
	
\subsection{Properties}
\label{ch13:sec:HRproperties}

As described in the previous section, for simple graph structures tailor-made and often non-parametric methods yield efficient statistical inference. For general graphs, a suitable parametric model class is typically required to specify arbitrary conditional independence structures and to perform parameter estimation and structure learning.    
For extremal graphical models, the \index{H\"usler--Reiss distribution}\HR{} distribution \citep{hueslerReiss1989} constitutes the only parametric family with the desired properties. Most importantly, in this class, extremal conditional independence is encoded in a parametric way, namely as the zero pattern of a certain \HR{} \index{precision matrix}precision matrix. This is one of the reasons why this distribution can be seen as an analogue of the Gaussian distribution in extremes.  

Recall from \cref{ex:HR} that an extreme value model with \HR{} distribution is parameterized by a variogram matrix $\Gamma$  and defined through the density of its exponent measure. 
The matrix $\Theta$ that appears in this density is called the \HR{} precision matrix and plays a crucial role for graphical modeling. It is in one-to-one correspondence with the variogram matrix through the mapping $\Theta = (P(-\Gamma/2)P)^+$, which is homeomorphic between $\C^d$ and the set $\Sone$ of symmetric, positive semidefinite matrices with zero row sums and rank $\d-1$ \citep{hentschelEngelkeSegers2022}.

The \HR{} distribution is widely used in applications due to its flexibility for statistical modeling \citep{eng_fondeville_oesting,deFondeville2018,thibaud_davison}. It turns out to be the only class of exponent measure densities that exhibits the structure of a pairwise interaction model \citep{lalancette2023pairwise}, an interpretable and computationally advantageous exponential family that is ubiquitous in high dimensional dependence and graphical modeling \citep{YRAL15,LDS16,KOBMK20}. Besides these properties, the main importance of the \HR{} family stems from the fact that extremal conditional independence and the extremal graphical structure can directly be read off from the precision matrix as
\[
\exCondInd{i}{j}{\V \setminus \{i, j\} \quad \Leftrightarrow \quad \Theta_{ij} = 0};
\]
\citep{engelkeHitz2020,hentschelEngelkeSegers2022}. This fact is the basis for virtually all estimation and structure learning algorithms that we present in the sequel.

\begin{figure}
	\centering
	\begin{subfigure}{0.3\textwidth}
		\includePlot[width=1\textwidth]{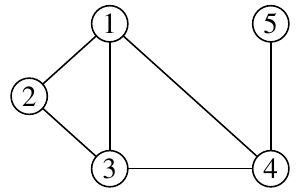}
		\vspace{-0.6cm}
	\end{subfigure}
	\begin{subfigure}{0.6\textwidth}
		\begin{align*}
			\left(\input{figures/G0_matrix.tex}\right)
			\mapsto
			\left(\input{figures/G1_matrix.tex}\right)
		\end{align*}
	\end{subfigure}
	\caption{
		Example of a decomposable graph and an associated partial and completed extremal variogram. Underlined entries, corresponding to non-edges, are uniquely recovered from the known variogram entries on edges. Note that unlike in the case of a block graph (\cref{ch13:fig:GammaCompletionBlock}), the solution is not simply obtained by summing the entries along paths. For example, $\Gamma_{24} = 15$ is neither equal to $\Gamma_{21} + \Gamma_{14} = 13$, nor to $\Gamma_{23} + \Gamma_{34} = 21$.
	}
	\label{ch13:fig:GammaCompletion}
\end{figure}

For \HR{} tree models, it is enough to specify the parameter matrix $\Gamma$ on the edges of the graph and the remaining entries are implied by the graphical structure through property~\eqref{tree_metric}.
This result can in fact be generalized to arbitrary graphs $\G = (\V,\E)$. 
Let $\mathring\Gamma$ be a partially specified matrix with given entries for $(i,j) \in \E$ such that all (fully specified) sub-matrices corresponding to cliques are valid variograms. The matrix completion problem of finding a valid variogram matrix $\Gamma \in \C^d$ with corresponding \HR{} precision matrix $\Theta$ such that   
\begin{align}\label{completion}
	\begin{cases}
	\Gamma_{ij}=\mathring{\Gamma}_{ij} & (i,j) \in \E,\\
	{\Theta}=0 & (i,j)\not\in \E,\\
\end{cases}
\end{align}
then has a unique solution \citep{hentschelEngelkeSegers2022}, where, if the graph is non-decomposable, we need the additional assumption that any valid, possibly non-graph structured solution exists. In practice, the solution $\Gamma$ is obtained through a recursive algorithm, which for decomposable graphs finishes after finitely many steps. Figure~\ref{ch13:fig:GammaCompletion} illustrate the matrix completion problem and its solution on an example of a decomposable graph.

It is worthwhile to highlight some similarities and difference of \HR{} and Gaussian distributions. At first glance, \eqref{eq:HRdensityGamma} appears to be the density of a multivariate lognormal distribution.
A crucial distinction is in that the precision matrix $\Theta$ is rank deficient, and in the ``direction'' of the vector $\bone$, the density decays at a Pareto rather than lognormal rate. This ensures homogeneity of the exponent measure density $\emd$, and also explains why it is not integrable on $\EE$.
To make the connection to Gaussian distributions more explicit, we consider the multivariate Pareto distribution $\Y$ that corresponds to a \HR{} model.
On the log-scale, it admits the stochastic representation
\begin{align}	
	\log\Y \mid \{\bone^\top \log\Y > 0\} = R\bone + \W,	\label{eq:HR_stochrepW}
\end{align}
where $R$ is a standard exponential random variable and, independently, $\W$ has a degenerate Gaussian distribution with covariance matrix $\Sigma=\Theta^+$.
Compared to classical Gaussian graphical models, the main difficulty of working with \HR{} graphical models, both in terms of theory and statistical methodology, is the zero row sums property on the precision matrix $\Theta$, as well as its rank which is constrained to be $d-1$.

\subsection{Parameter estimation}
\label{ch13:sec:parameterEstimation}

\subsubsection{Maximum likelihood on known graph}

Similarly to \cref{ch13:sec_para_est_simple}, suppose that we have $n$ independent copies of a random vector $\X$ in the domain of attraction of a \HR{} distribution with parameter matrix $\Gamma$. In order to use properties of Gaussian \index{maximum likelihood estimation}maximum likelihood theory we use the transformation $\Ym$ of the corresponding multivariate Pareto distribution $\Y$, whose increments have a lognormal distribution with mean vector $-\diag(\Sigma^{(m)})/2$ and covariance matrix $\Sigma^{(m)}$; see \cref{Ym_HR}. Similarly to the empirical variogram estimator in~\eqref{emp_vario}, out of the $n$ samples, we choose the $k$ largest observations in the $m$th component and apply the log increments transformation in~\eqref{Y_normal}. The resulting dataset then has approximately the surrogate positive semidefinite Gaussian log-likelihood 
\begin{align}
	\ell(\Theta;\overline \Gamma) = \log\operatorname{Det}(\Theta) + \frac12 \tr(\Theta \overline \Gamma), \label{eq:surr_loglik}
\end{align}
which we parameterize in terms of \HR{} precision matrices $\Theta \in \Sone$, and $\overline \Gamma$ is an estimator of $\Gamma$; see \cite{REZ2023,hentschelEngelkeSegers2022} for a detailed derivation. 
It is a surrogate likelihood since the mean is empirically normalized to zero and $\overline \Gamma$ can be any estimator of $\Gamma$, and it is approximate since the exceedances of $\X$ only follow a multivariate Pareto distribution in the limit. Nevertheless, this likelihood turns out to be computationally tractable and accurate enough to provide estimators with good statistical properties. In order to use all data, the most common choice is to combine data from conditioning on all components $m\in\V$ and use the joint empirical variogram $\overline \Gamma = \hat \Gamma$ from~\eqref{emp_vario_joint}.






Suppose now that the \HR{} distribution is an extremal graphical model on the known graph $\G = (\V,\E)$. In order to obtain a first estimator that is consistent for the entries of $\Theta$ and has the zeros at the correct entries, we can maximize  the log-likelihood~\eqref{eq:surr_loglik} under the constraint that 
\begin{align}\label{theta_cons}
	 \Theta_{ij}=0 \quad (i,j)\not\in \E.
\end{align}
Surprisingly, the maximizer $\hat \Theta_{\G}$ of this optimization problem is identical to the solution of the completion problem~\eqref{completion} with partial variogram matrix chosen as $\mathring{\Gamma}_{ij} = \overline \Gamma$ for $(i,j) \in \E$.
In practice, we can therefore proceed as follows: choose an estimator of the matrix $\Gamma$ such as the joint empirical variogram $\hat \Gamma$; use the entries of $\hat \Gamma$ on the edges of $\G$ in the partial matrix $\mathring \Gamma$ as input for the matrix completion problem together with~\eqref{theta_cons}; apply the algorithm of \cite{hentschelEngelkeSegers2022} to obtain the completion $\hat \Gamma_{\G}$ and note that it exists and is unique since $\mathring \Gamma$ has the valid completion $\hat \Gamma$.
The resulting estimator $\hat \Gamma_{\G}$ is consistent and its precision $\hat \Theta_{\G}$ has extremal graph structure $\G$ \cite[Section 5.2]{hentschelEngelkeSegers2022}.

\subsubsection{Colored graphical models}\index{colored graphical model}

In addition to the graphical constraint~\eqref{theta_cons}, it might be desirable to further reduce the dimensionality of the parameter matrix.
A potential approach are parameter symmetries, which can be visualized through colored graphs \citep{roettger2023parametric}.
For an undirected graph $\G=(\V,\E)$ let $\lambda: \E \to \{1,\dots ,r\}$ be an edge coloring function that maps every edge to a natural number $1,\ldots,r$.
In inference, graph colorings can for example be determined using clustering algorithms.
A H\"usler--Reiss restricted concentration (RCON) model is a linear submodel of H\"usler--Reiss graphical models on $\G$ with additional symmetries
\[\Theta_{st}=\Theta_{uv} \text{ whenever } \lambda(st)=\lambda(uv) \text{ for } (s,t),(u,v)\in \E. \]
Such models allow for a simple surrogate maximum likelihood estimator, as all constraints are affine in the natural parameter of the underlying exponential family. 
Alternatively, one may impose the symmetries on the level of the variogram matrix $\Gamma$, resulting in the H\"usler--Reiss restricted variogram (RVAR) model with
\[\Gamma_{st}=\Gamma_{uv} \text{ whenever } \lambda(st)=\lambda(uv) \text{ for } (s,t),(u,v)\in \E. \]
The RVAR model usually does not allow an affine description in the precision matrix $\Theta$, such that simple surrogate maximum likelihood estimation may be difficult to apply.
As an alternative, \cite{roettger2023parametric} employ the two-step estimation procedure of \cite{LZ2022}, which allows to efficiently impose linear constraints in a mixed parametrization of an exponential family.
Depending on the number of edge color classes $r$, both RCON and RVAR models result in a drastic parameter reduction while showing good performance in applications \citep{roettger2023parametric}.

\subsubsection{Positive dependence}
\label{ch13:sec:structureLearning:emtp2}

\index{multivariate total positivity}Multivariate total positivity ($\text{MTP}_2$) and positive association are properties of a multivariate random vector that describe more or less strong   notions of positive dependence. Many datasets naturally exhibit an intrinsic positivity and statistical methodology based on this notion can therefore improve accuracy and interpretability; see \cite{LZ2022} and references therein.


For multivariate Pareto distributions, \cite{REZ2023} introduce extremal \MTPtwo{} (\EMTPtwo{}) and extremal association by requiring that for all $m\in \V$ the vectors $\Ym$ in~\eqref{Ym} be \MTPtwo{} and associated in the usual sense, respectively.
For \HR{} distributions, \EMTPtwo{} is equivalent to the parametric constraint that the parameter matrix $\Theta$ is a graph Laplacian, that is,
\begin{align}\label{emtp2}
	\Theta_{ij}\le 0\quad \forall i\neq j.
\end{align}
In particular, this implies that all \HR{} tree models are \EMTPtwo{}.  
To estimate a \HR{} model that is \EMTPtwo{}, \cite{REZ2023} propose to maximize likelihood~\eqref{eq:surr_loglik} under the linear constraint~\eqref{emtp2}. The resulting estimator $\hat \Theta_{+}$  typically contains many zeros, without enforcing this explicitly in the optimization problem.
If the underlying \HR{} distribution is \EMTPtwo{} and an extremal graphical model on $\G =(\V, \E)$, then the estimated graph $\G_+ = (\V, \hat \E_+)$  corresponding to $\hat \Theta_{+}$ is asymptotically a supergraph of $\G$, that is, $\PP(\E\subseteq\hat \E_+)\to 1$ as the sample size $n\to\infty$.


A sufficient condition for extremal association for H\"usler--Reiss distributions is the metric property, which requires the triangle inequalities
\[\Gamma_{ij}\le \Gamma_{ik}+\Gamma_{jk} \]
to hold for any triple $i,j,k\in \V$. An \EMTPtwo{} \HR{} distribution always satisfies this property, while the inverse is only true for $\d=3$.
While \EMTPtwo{} is by construction a global property, it can be a too strong assumption for applications with localized positive dependence structure.
For a given \HR{} graphical model on $\G$, \cite{roettgerSchmitz2023} therefore propose a model with local metric property where the triangle inequalities are only required for all triples $i,j,k\in C$, for all cliques $C$ of the graph $\G$. They estimate this model based on mixed convex exponential families \citep{LZ2022} in a two-step procedure.

\subsubsection{Score matching}
\label{ch13:sec:structureLearning:score}
\index{score matching}Score matching is an alternative to likelihood inference that proposes to minimize the Fisher information distance \citep{Hyvaerinen2005,drt2017}.
The approach was originally proposed because its loss function does not depend on the normalizing constant of the underlying distribution, but also became popular for multivariate Gaussian models for its attractive computational properties.

In graphical extremes, \cite{lederer2023extremes} propose a computationally efficient score matching estimator for a family of distributions that generalizes the \HR{} model.
Specifically, they introduce a class of pseudo-densities that is parameterized by a superset of all \HR{} precision matrices and a nuisance location parameter. This function class contains all \HR{} multivariate Pareto densities, but also improper and proper (though not multivariate Pareto) densities. Since the score matching objective is well-defined even for non-integrable densities, they are able to define an $L^1$ penalized score matching estimator in this class.
A main contribution in \cite{lederer2023extremes} is in providing an estimator of the \HR{} precision matrix that, despite not always being a valid \HR{} precision matrix itself, enjoys sharp concentration guarantees in high dimensions. Nevertheless, the regularization scheme inherently produces sparsity in the estimated precision matrix.

\subsection{Structure learning}
\label{ch13:sec:structureLearning}

\subsubsection{An extremal graphical lasso}

The representation of conditional independence as zeros in the precision matrix and the surrogate log-likelihood in~\eqref{eq:surr_loglik} suggest the use of $L^1$ penalization methods to enforce graph sparsity. The Gaussian \index{graphical lasso}graphical lasso \citep{yuan2007model,banerjee2008,friedmanEtAl2007} could then be adapted to estimate sparse H\"usler--Reiss graphical models, giving rise to an extremal graphical lasso problem of the form
\begin{equation} \label{eq:eglasso}
	\minimize_{\Theta \in \SHR} -\ell(\Theta;\overline{\Gamma}) + \lambda \onenormoff{\Theta},
\end{equation}
where $\onenormoff{\Theta} := \mathop{\sum\sum}_{i \neq j} |\Theta_{ij}|$ and $\lambda \geq 0$ is a penalty parameter that controls the level of sparsity in the solution. The matrix $\overline \Gamma$ can be a general estimator of $\Gamma$ but will typically be chosen as the joint empirical variogram $\hat \Gamma$. The Gaussian graphical lasso has implementations that are fast and stable in high dimensions, and simultaneously learns a graph structure and yields a model estimate on the learned graph. An estimator obtained by solving \eqref{eq:eglasso} would share this property. However, an issue with this optimization problem is the positive semi-definiteness of the \HR{} precision matrix. Indeed, the interaction between the $L^1$ penalization and constraint of zero row sums in $\Theta$ has been shown to lead to dense estimated models where every parameter is estimated to be nonzero \citep{ying2020does,ying2021minimax}.
We describe two approaches that circumvent this issue, and which can be regarded as approximations of the ideal extremal graphical lasso problem in \eqref{eq:eglasso}. 

\subsubsection{Graph learning through majority voting}
\label{ch13:sec:structureLearning:eglearn}

In \cite{engelke2022a}, the authors consider the matrices $\Theta^{(1)}, \dots, \Theta^{(d)} \in \RR^{(d-1) \times (d-1)}$, indexed by $\V\setminus \{m\}$, where $\Theta^{(m)}$ is obtained by removing the $m$th row and column of $\Theta$; these are the inverses of the covariance matrices in~\eqref{Y_normal}. 
The authors leverage the fact that for each $m$, $\Theta^{(m)}$ parameterizes the \HR{} model and encodes most of the extremal graphical structure in its sparsity pattern: by the properties of $\Theta$, for any $i,j \neq m$,
\[
	\exCondInd{i}{j}{\V \setminus \{i, j\}} \quad \Leftrightarrow \quad \Theta_{ij}^{(m)} = 0;
\]
see \cite[Section 4.3]{engelkeHitz2020}. Conditional independencies involving node $m$ are instead encoded as zero row sums in $\Theta^{(m)}$. These matrices are usual positive definite precision matrices, and thus, their support can be estimated by the Gaussian graphical lasso, neighborhood selection \citep{meinshausen2006} or similar algorithms. The estimated support of $\Theta^{(m)}$ offers information on the presence of any possible edge not connected to node $m$ in the graph $\G$. By aggregating the estimated supports of $\Theta^{(1)}, \dots, \Theta^{(d)}$ through the majority voting meta-algorithm \eglearn{}, \cite{engelke2022a} obtain a full estimated extremal graph $\hat \G$. The method is moreover shown to be model selection consistent (sparsistent) in exponentially growing dimension under usual conditions. \eglearn{} is a pure structure learning method. In combination with the matrix completion of the empirical variogram $\hat \Gamma$ on the estimated graph $\hat \G$ in a second step, the method provides a consistent and sparsistent estimator of the \HR{} parameter matrix~\citep{hentschelEngelkeSegers2022}.

\subsubsection{Graph learning through parameter shift}
\label{ch13:sec:structureLearning:shift}

In \cite{wan2023graphical}, it is proposed to estimate the shifted parameter matrix $\Theta^* := \Theta + c\bone\bone^\top$, for an explicit positive constant $c$. As opposed to $\Theta$, $\Theta^*$ is a precision matrix of full rank and its inverse $\Sigma^*$ can be estimated from data. The authors then solve the graphical lasso type optimization problem
\[
	\minimize_{\Theta^*} -\log\det(\Theta^*) + \tr(\Theta^* \hat\Sigma^*) + \lambda \onenormoff{\Theta^* - c\bone\bone^\top},
\]
where $\Theta^*$ ranges over the symmetric positive definite matrices and where $\hat\Sigma^*$ is an estimator of $\Sigma^*$. The idea is that the first two terms in this objective function can be seen as a negative Gaussian pseudolikelihood, which admits a minimum close to $\Theta^*$ if the estimator $\hat\Sigma^*$ is sufficiently precise. The nonstandard penalty term enforces sparsity in $\Theta$ rather than in $\Theta^*$. The method is fast and has the advantage of simultaneously learning a graph and producing an estimator of $\Theta$. However, the estimator is not guaranteed to be a valid H\"usler--Reiss precision matrix. Nevertheless, partially relying on concentration results proved in \cite{engelke2022a}, \cite{wan2023graphical} establish consistency in terms of both model selection and parameter estimation. 

Note that both this method and \eglearn{} estimate the sparsity pattern of surrogate, full rank precision matrices. By doing so, they avoid the ill behavior of positive semi-definite $L^1$ penalized maximum likelihood estimation. They sacrifice, however, the special structure of the zero row sums of \HR{} precision matrices. For this reason, they share an important shortcoming: neither is guaranteed to produce a connected estimated graph, despite the fact \HR{} graphical models are by definition connected. Solving the ideal extremal graphical lasso problem in \eqref{eq:eglasso} would remedy this issue, in addition to producing a valid parameter estimate $\hat\Theta \in \SHR$.


	\section{Application}
	\label{ch13:sec:application}

\newcommand{\valFlightsTexasDataSetName}{Texas Flights Data}
\newcommand{\valFlightsTexasDataSetNiceName}{Texas Flights Data}
\newcommand{\valFlightsTexasUsedP}{0.95}
\newcommand{\valFlightsTexasIndexEstStart}{2005-01-01}
\newcommand{\valFlightsTexasIndexEstEnd}{2010-12-31}
\newcommand{\valFlightsTexasIndexValStart}{2011-01-01}
\newcommand{\valFlightsTexasIndexValEnd}{2020-12-31}
\newcommand{\valFlightsTexasNObsEst}{1764}
\newcommand{\valFlightsTexasNObsVal}{1839}
\newcommand{\valFlightsTexasNVertices}{29}
\newcommand{\valFlightsTexasExpertName}{Flight Graph}
\newcommand{\valFlightsTexasNObsOriginal}{6117}
\newcommand{\valFlightsTexasNVerticesOriginal}{420}
\newcommand{\valFlightsTexasNObs}{3603}
\newcommand{\valFlightsTexasIndexStart}{2005-01-01}
\newcommand{\valFlightsTexasIndexEnd}{2020-12-31}
\newcommand{\valDanubeDataSetName}{Danube Flow Data}
\newcommand{\valDanubeDataSetNiceName}{Danube Flow Data}
\newcommand{\valDanubeUsedP}{0.9}
\newcommand{\valDanubeIndexEstStart}{1960}
\newcommand{\valDanubeIndexEstEnd}{2010}
\newcommand{\valDanubeIndexValStart}{1960}
\newcommand{\valDanubeIndexValEnd}{2010}
\newcommand{\valDanubeNObsEst}{214}
\newcommand{\valDanubeNObsVal}{214}
\newcommand{\valDanubeNVertices}{31}
\newcommand{\valDanubeExpertName}{Flow Graph}
\newcommand{\valDanubeNObs}{428}
\newcommand{\valDanubeIndexStart}{1960}
\newcommand{\valDanubeIndexEnd}{2010}
\newcommand{\valExpertKnowledgeMethodName}{Expert Knowledge}
\newcommand{\valExpertKnowledgeMethodShortName}{expertKnowledge}
\newcommand{\valFullVarioMethodName}{Full Variogram}
\newcommand{\valFullVarioMethodShortName}{fullVario}
\newcommand{\valRandomGraphMethodName}{Random Graph}
\newcommand{\valRandomGraphMethodShortName}{randomGraph}
\newcommand{\valEmstMethodName}{EMST}
\newcommand{\valEmstMethodShortName}{emst}
\newcommand{\valEglearnMethodName}{\eglearn{}}
\newcommand{\valEglearnMethodShortName}{eglearn}
\newcommand{\valEmtpMethodName}{\EMTPtwo{}}
\newcommand{\valEmtpMethodShortName}{emtp}
\newcommand{\valWazMethodName}{Parameter Shift}
\newcommand{\valWazMethodShortName}{waz}
\newcommand{\valColorsMethodName}{Colored Graph}
\newcommand{\valColorsMethodShortName}{colors}
\newcommand{\valLeoMethodName}{Score Matching}
\newcommand{\valLeoMethodShortName}{leo}
\newcommand{\valFlightsTexasExpertKnowledgeNEdges}{129}
\newcommand{\valFlightsTexasExpertKnowledgeLoglik}{-9.23}
\newcommand{\valFlightsTexasExpertKnowledgeAIC}{276.46}
\newcommand{\valFlightsTexasExpertKnowledgeBIC}{876.34}
\newcommand{\valFlightsTexasFullVarioNEdges}{406}
\newcommand{\valFlightsTexasFullVarioLoglik}{-293.68}
\newcommand{\valFlightsTexasFullVarioAIC}{1399.37}
\newcommand{\valFlightsTexasFullVarioBIC}{3287.38}
\newcommand{\valFlightsTexasRandomGraphNEdges}{100}
\newcommand{\valFlightsTexasRandomGraphLoglik}{-713.79}
\newcommand{\valFlightsTexasRandomGraphAIC}{1627.58}
\newcommand{\valFlightsTexasRandomGraphBIC}{2092.61}
\newcommand{\valFlightsTexasEmstNEdges}{28}
\newcommand{\valFlightsTexasEmstLoglik}{-6435.2}
\newcommand{\valFlightsTexasEmstAIC}{12926.41}
\newcommand{\valFlightsTexasEmstBIC}{13056.61}
\newcommand{\valFlightsTexasEglearnNEdges}{101}
\newcommand{\valFlightsTexasEglearnLoglik}{1123.45}
\newcommand{\valFlightsTexasEglearnAIC}{-2044.9}
\newcommand{\valFlightsTexasEglearnBIC}{-1575.22}
\newcommand{\valFlightsTexasEmtpNEdges}{126}
\newcommand{\valFlightsTexasEmtpLoglik}{1212.97}
\newcommand{\valFlightsTexasEmtpAIC}{-2173.94}
\newcommand{\valFlightsTexasEmtpBIC}{-1588.01}
\newcommand{\valFlightsTexasWazNEdges}{142}
\newcommand{\valFlightsTexasWazLoglik}{1192.02}
\newcommand{\valFlightsTexasWazAIC}{-2100.04}
\newcommand{\valFlightsTexasWazBIC}{-1439.7}
\newcommand{\valFlightsTexasColorsNEdges}{126}
\newcommand{\valFlightsTexasColorsLoglik}{1341.17}
\newcommand{\valFlightsTexasColorsAIC}{-2430.34}
\newcommand{\valFlightsTexasColorsBIC}{-1844.4}
\newcommand{\valFlightsTexasLeoNEdges}{173}
\newcommand{\valFlightsTexasLeoLoglik}{1144.97}
\newcommand{\valFlightsTexasLeoAIC}{-1943.95}
\newcommand{\valFlightsTexasLeoBIC}{-1139.45}
\newcommand{\valDanubeExpertKnowledgeNEdges}{30}
\newcommand{\valDanubeExpertKnowledgeLoglik}{-430.82}
\newcommand{\valDanubeExpertKnowledgeAIC}{921.65}
\newcommand{\valDanubeExpertKnowledgeBIC}{982.94}
\newcommand{\valDanubeFullVarioNEdges}{465}
\newcommand{\valDanubeFullVarioLoglik}{-2648.5}
\newcommand{\valDanubeFullVarioAIC}{6227}
\newcommand{\valDanubeFullVarioBIC}{7177.02}
\newcommand{\valDanubeRandomGraphNEdges}{60}
\newcommand{\valDanubeRandomGraphLoglik}{-1849.19}
\newcommand{\valDanubeRandomGraphAIC}{3818.38}
\newcommand{\valDanubeRandomGraphBIC}{3940.97}
\newcommand{\valDanubeEmstNEdges}{30}
\newcommand{\valDanubeEmstLoglik}{-618.44}
\newcommand{\valDanubeEmstAIC}{1296.89}
\newcommand{\valDanubeEmstBIC}{1358.18}
\newcommand{\valDanubeEglearnNEdges}{82}
\newcommand{\valDanubeEglearnLoglik}{-500.91}
\newcommand{\valDanubeEglearnAIC}{1165.81}
\newcommand{\valDanubeEglearnBIC}{1333.34}
\newcommand{\valDanubeEmtpNEdges}{72}
\newcommand{\valDanubeEmtpLoglik}{-489.62}
\newcommand{\valDanubeEmtpAIC}{1123.24}
\newcommand{\valDanubeEmtpBIC}{1270.34}
\newcommand{\valDanubeWazNEdges}{146}
\newcommand{\valDanubeWazLoglik}{-583.35}
\newcommand{\valDanubeWazAIC}{1458.69}
\newcommand{\valDanubeWazBIC}{1756.98}

\subsection{Dataset}

Applications of extremal graphical models range from modeling river flows for flood risk assessment \citep{asadi2015extremes,engelkeHitz2020,asenova2021inference,wan2023graphical}, over financial applications \citep{klu2021scaling, engelkeVolgushev2022} to food nutrition \citep{buc2021,krali2023heavy}.

In this section we illustrate the methods presented in
\cref{ch13:sec:hueslerReiss,ch13:sec:simpleGraphStructures}
by applying them to the flight delay dataset introduced in \cite{hentschelEngelkeSegers2022},
which is available from the \texttt{R} package \texttt{graphicalExtremes} \citep{graphicalExtremes2022}. The goal is to qualitatively compare the methods and to illustrate their specific properties.    
The dataset contains daily accumulated (positive) flight delays in minutes
at major US airports from 2005 to 2020.
We consider the airports in the ``Texas Cluster''
obtained in \cite{hentschelEngelkeSegers2022} as the result of a $k$-medoids clustering algorithm.
The resulting dataset consists of $n=$ \valFlightsTexasNObs{} observations at $d=$ \valFlightsTexasNVertices{} airports. It can be obtained as follows:
\begin{Verbatim}[samepage=true]
library(graphicalExtremes)
delays <- getFlightDelayData(airportFilter = 'tcCluster',
                             dateFilter = 'tcAll')
\end{Verbatim}
The airports and the graph that contains an edge whenever there is at least one monthly flight between two airports is shown in \cref{ch13:fig:FlightsTexasGraphsExpertKnowledge}.
We split the data into a training set
(\valFlightsTexasIndexEstStart{} to \valFlightsTexasIndexEstEnd)
and a test set
(\valFlightsTexasIndexValStart{} to \valFlightsTexasIndexValEnd).
The former is used to perform structure learning and parameter estimation with the different methods and possibly different hyperparameters.
The latter is used to compare the fitted models in terms of the test likelihoods and other summary statistics.


\subsection{Methods}


Throughout, we assume a \HR{} distribution to model the extremal dependence structure of this dataset. We choose the empirical extremal variogram $\hat \Gamma$ defined in~\eqref{emp_vario} as input for all methods that we apply.
Recall that the marginal distributions are empirically normalized to standard Pareto distributions; see \cref{ch13:sec_para_est_simple}.
As a threshold we use the empirical marginal $p = 1 - k/n = 95\%$ quantiles so that the effective sample size in each direction is $k=88$ in the training set.
We use the \HR{} model with parameter matrix given by the empirical variogram as a first benchmark and note that it corresponds to a fully connected extremal graph. To obtain two other benchmarks with non-trivial graph structures, we choose the ``Flight Graph'' obtained by considering all connections with at least one monthly flight, and a randomly generated graph structure. For both graphs, we fit a \HR{} model by completing the empirical variogram using the matrix completion in \cref{ch13:sec:parameterEstimation}.

The ``\valEglearnMethodName{}''
method in \cref{ch13:sec:structureLearning:eglearn} and the extremal minimum spanning tree ``\valEmstMethodName{}'' in \cref{ch13:sec_structure_simple} only provide an estimator for the underlying extremal graphical structure $\G$, but not for the parameter matrix $\Gamma$. In this application, the estimated parameter matrix of the ``\valWazMethodName{}'' method in \cref{ch13:sec:structureLearning:shift} does not return a valid $\Gamma$ matrix, and we therefore only use the estimated graph.
In the case of ``\valEglearnMethodName{}'', the algorithm used to estimate the support of the precision matrices $\Theta^{(m)}$ is neighborhood selection with common penalty parameter $\rho$ \citep{engelke2022a}.
For these three methods, we obtain a full  
model by completing the empirical variogram on the estimated graphs as explained in \cref{ch13:sec:parameterEstimation}. 

The  ``\valEmtpMethodName{}'' method in \cref{ch13:sec:structureLearning:emtp2} and the ``\valLeoMethodName{}'' method in \cref{ch13:sec:structureLearning:score} directly yield estimates of the parameter matrix $\Gamma$ and a graph structure $\G$.
The method ``\valColorsMethodName{}'' fits an RVAR model where we choose the ``\valEmtpMethodName{}'' estimate as the underlying graph and use a clustering algorithm to assign edge colorings. 

Some of above methods are available in the R package \texttt{graphicalExtremes}:
\begin{Verbatim}[samepage=true]
p <- 0.95
G_hat <- emp_vario(delays, p = p)
models_eglearn <- eglearn(delays, p)
model_emst <- emst(delays, p)
G_empt2 <- emtp2(G_hat)
\end{Verbatim}

\subsection{Results}

\begin{figure}[!ht]
    \newcommand{\sfheight}{5cm}
    \centering
    \begin{subfigure}{0.34\textwidth}
        \includePlot[height=\sfheight]{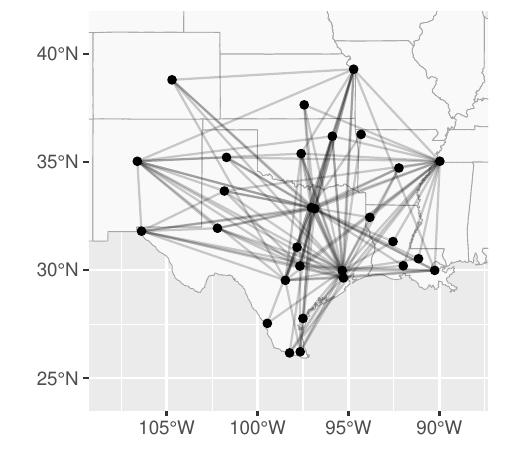}
        \subcaption{\valFlightsTexasExpertName}
        \label{ch13:fig:FlightsTexasGraphsExpertKnowledge}
    \end{subfigure}
    \begin{subfigure}{0.3\textwidth}
        \includePlot[height=\sfheight]{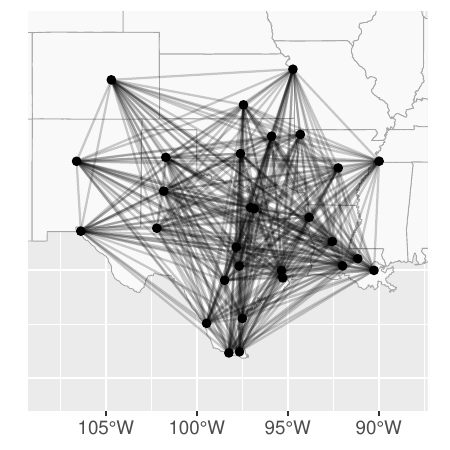}
        \subcaption{\valFullVarioMethodName}
        \label{ch13:fig:FlightsTexasGraphsFullVario}
    \end{subfigure}
    \begin{subfigure}{0.3\textwidth}
        \includePlot[height=\sfheight]{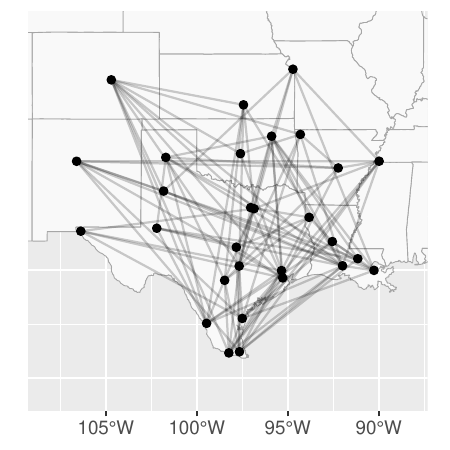}
        \subcaption{\valRandomGraphMethodName}
        \label{ch13:fig:FlightsTexasGraphsRandomGraph}
    \end{subfigure}

    \begin{subfigure}{0.34\textwidth}
        \includePlot[height=\sfheight]{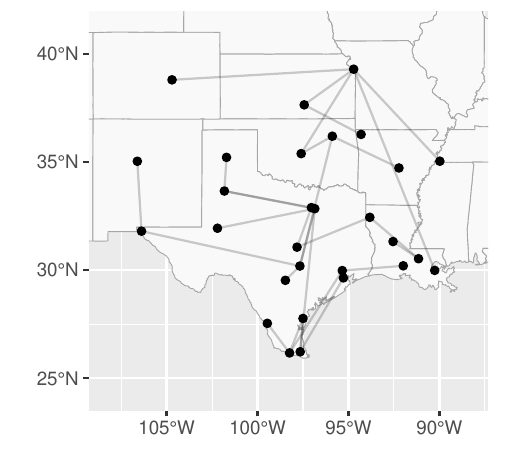}
        \subcaption{\valEmstMethodName}
        \label{ch13:fig:FlightsTexasGraphsEmst}
    \end{subfigure}
    \begin{subfigure}{0.3\textwidth}
        \includePlot[height=\sfheight]{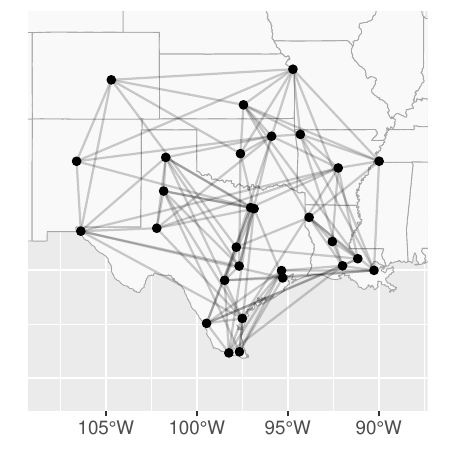}
        \subcaption{\valEglearnMethodName}
        \label{ch13:fig:FlightsTexasGraphsEglearn}
    \end{subfigure}
    \begin{subfigure}{0.3\textwidth}
        \includePlot[height=\sfheight]{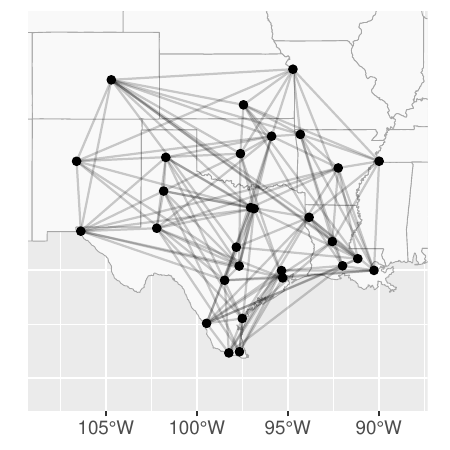}
        \subcaption{\valEmtpMethodName}
        \label{ch13:fig:FlightsTexasGraphsEmtp}
    \end{subfigure}

    \begin{subfigure}{0.34\textwidth}
        \includePlot[height=\sfheight]{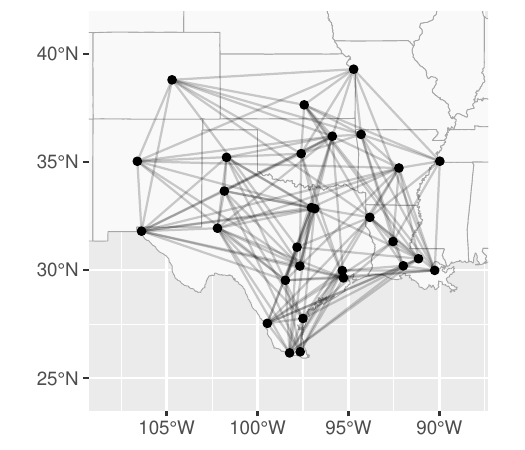}
        \subcaption{\valWazMethodName}
        \label{ch13:fig:FlightsTexasGraphsWaz}
    \end{subfigure}
    \begin{subfigure}{0.3\textwidth}
        \includePlot[height=\sfheight]{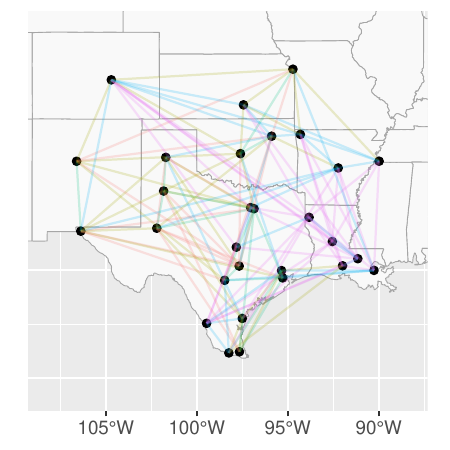}
        \subcaption{\valColorsMethodName}
        \label{ch13:fig:FlightsTexasGraphsColors}
    \end{subfigure}
    \begin{subfigure}{0.3\textwidth}
        \includePlot[height=\sfheight]{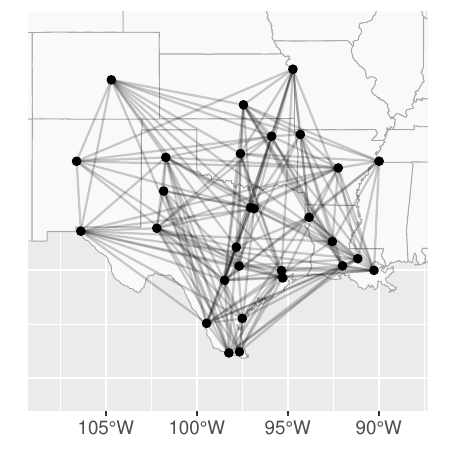}
        \subcaption{\valLeoMethodName}
        \label{ch13:fig:FlightsTexasGraphsLeo}
    \end{subfigure}
    \caption{
        Graph structures estimated by the different methods.
        In (\subref{ch13:fig:FlightsTexasGraphsColors}),
        edge colors correspond to variogram entries with the same value.
    }
    \label{ch13:fig:FlightsTexasGraphs}
\end{figure}

\cref{ch13:fig:FlightsTexasGraphs}
shows the graph structures estimated by the different methods
and \cref{ch13:table:ResultsFlightsTexas} contains the corresponding edge counts and log-likelihoods on the test data. 
\cref{ch13:fig:FlightsTexasParams}
compares the variogram entries implied by the fitted models
to the corresponding entries of the empirical variograms on the test data.

By construction, the sparsest estimated graph is the ``\valEmstMethodName{}'' estimate. On this data, the test likelihood and the consistent underestimation of the non-adjacent empirical variograms indicate that a denser graph is necessary to capture the full dependence structure. On the other side of the spectrum is the full empirical variogram corresponding to a dense, fully connected graph. The test likelihood suggest that this model is not sparse enough and overfits to the training data. The random graph with an ad-hoc number of 100 edges is in between the two methods in terms of test likelihood. The additional edges compared to the tree allow for more flexible modeling, but since the graph is not learned in a data-driven way, it is not competitive with the other methods. The ``\valFlightsTexasExpertName{}'' uses a similarly sparse graph, but the edges are chosen in a more reasonable way according to the existence of flight connections. This results in a better test likelihood than the previous methods.

\begin{figure}[!ht]
    \newcommand{\sfheight}{5cm}
    \centering
    \begin{subfigure}{0.35\textwidth}
        \includePlot[height=\sfheight]{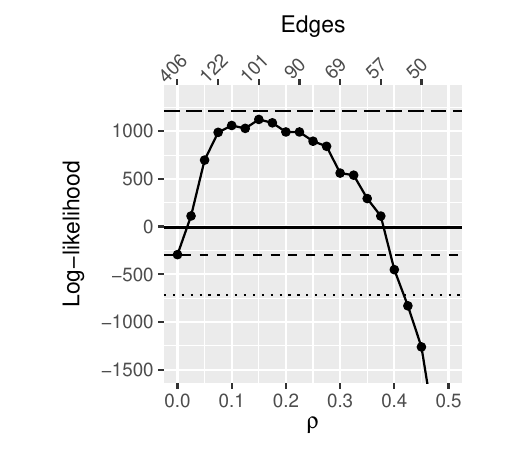}
        \subcaption{}
        \label{ch13:fig:FlightsTexasLogliksEglearn}
    \end{subfigure}%
    \hspace{-0.09\textwidth}
    \begin{subfigure}{0.3\textwidth}
        \includePlot[height=\sfheight]{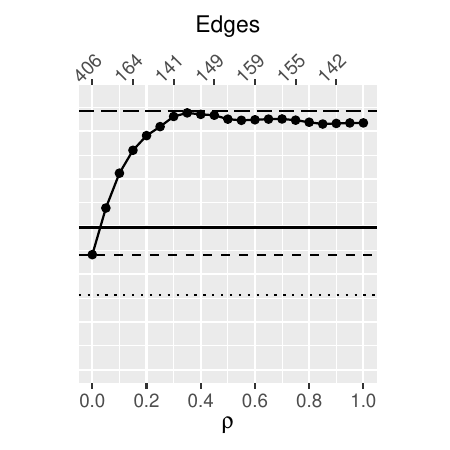}
        \subcaption{}
        \label{ch13:fig:FlightsTexasLogliksWaz}
    \end{subfigure}%
    \hspace{-0.1\textwidth}
    \begin{subfigure}{0.3\textwidth}
        \includePlot[height=\sfheight]{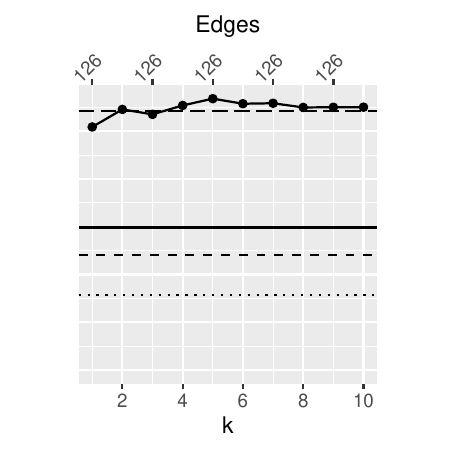}
        \subcaption{}
        \label{ch13:fig:FlightsTexasLogliksColors}
    \end{subfigure}%
    \hspace{-0.1\textwidth}
    \begin{subfigure}{0.3\textwidth}
        \includePlot[height=\sfheight]{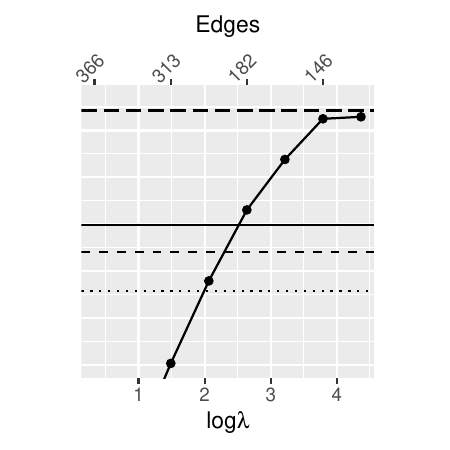}
        \subcaption{}
        \label{ch13:fig:FlightsTexasLogliksLeo}
    \end{subfigure}%
    \caption{%
        Test log-likelihood for different tuning parameters
        for
        (\subref{ch13:fig:FlightsTexasLogliksEglearn})~``\valEglearnMethodName{}'',
        (\subref{ch13:fig:FlightsTexasLogliksWaz})~``\valWazMethodName{}'',
        (\subref{ch13:fig:FlightsTexasLogliksColors})~``\valColorsMethodName{}'',
        and
        (\subref{ch13:fig:FlightsTexasLogliksLeo})~``\valLeoMethodName{}''.
        Methods without tuning parameters are indicated by horizontal lines
        (%
            ``\valRandomGraphMethodName{}'' dotted,
            ``\valFullVarioMethodName{}'' short-dashed,
            ``\valDanubeExpertName{}'' solid,
            ``\valEmtpMethodName{}'' long-dashed%
        ),
        ``\valEmstMethodName{}'' is below the axis and not shown.
    }
    \label{ch13:fig:FlightsTexasLogliks}
\end{figure}

The remaining methods, which all enforce sparsity in the extremal graph in an automatic, data-driven way, perform significantly better on this data set.
The methods ``\valLeoMethodName{}'', ``\valEglearnMethodName{}'', and ``\valWazMethodName{}'' each
have a penalty parameter,
and ``\valColorsMethodName{}'' has the number of colors as hyperparameter.
These methods were fitted for a range of different tuning parameters and the best-performing
model was chosen by considering the log-likelihood on the test data, as shown in \cref{ch13:fig:FlightsTexasLogliks}.
Furthermore, \cref{ch13:fig:FlightsTexasThetaConvergence} shows the convergence
of entries in $\Theta$ to zero as the penalization parameter is increased in the three structure learning methods,
illustrating their similarity to the classical graphical lasso.
Overall, the ``\valColorsMethodName{}'' seems to have a slightly better test likelihood than the other methods on this data set. Interestingly, this is not reflected in the bivariate summary statistics in \cref{ch13:fig:FlightsTexasParams}, which would hint at a slight overestimation of extremal dependence.



\begin{table}
    \centering
    \begin{tabular}{rrr}
  \hline
 & Edges & Log-likelihood \\ 
  \hline
Flight Graph & 129 & -9.23 \\ 
  Full Variogram & 406 & -293.68 \\ 
  Random Graph & 100 & -713.79 \\ 
  EMST & 28 & -6435.20 \\ 
  \eglearn{} & 101 & 1123.45 \\ 
  \EMTPtwo{} & 126 & 1212.97 \\ 
  Parameter Shift & 142 & 1192.02 \\ 
  Colored Graph & 126 & 1341.17 \\ 
  Score Matching & 173 & 1144.97 \\ 
   \hline
\end{tabular}

    \caption{
        Edge counts and test log-likelihoods for the different methods.
    }
    \label{ch13:table:ResultsFlightsTexas}
\end{table}

	\appendix
	\newpage
	\section{\texorpdfstring{$d$}{d}-separation}    
	\label{ch13:sec:appendixA}
	
Let $\G = (\V, \E)$ be a DAG.
The set of parents $\pa(i)$ of node $i$ is defined as all nodes $j$ such that $(j,i) \in \E$.	
The set of descendants $\des(i)$ of some node $i\in \V$ contains all nodes $j\in \V$ for which the graph contains a directed path $(i,\ldots,j)\in \mathcal{E}$.

\begin{definition}[Pearl's $d$-separation]\label{def:sep_dir}\index{$d$-separation}
	Let $\G=(\V,\E)$ be a directed acyclic graph.
	A path $(i_1,\ldots,i_m)$ is blocked by a set $C$ when it contains a node $i_k$ such that one of the following holds:
	\begin{itemize}
		\item $i_k\in C$ and $i_{k-1}\rightarrow i_k\rightarrow i_{k+1}$ or $i_{k-1}\leftarrow i_k\leftarrow i_{k+1}$ or $i_{k-1}\leftarrow i_k\rightarrow i_{k+1}$,
		\item $i_k\not\in C$ and $\des(i_k)\cap S=\emptyset$ and $i_{k-1}\rightarrow i_k\leftarrow i_{k+1}$. 
	\end{itemize}
	Disjoint sets $A,B\subseteq\V$ are called $d$-separated by a disjoint set $C$ when every path between nodes in $A$ and $B$ is blocked by $C$.
\end{definition}

Note that there are equivalent ways to define $d$-separation that rely on the notion of collider.
A collider is a vertex $i \in \V$ that has at least two parents. The moralized graph of $\G$ is defined by adding an edge, if not already present, between any two common parents of every collider (the orientation of that new edge is irrelevant). The undirected skeleton of this moralized graph is then obtained by remove the orientation of every edge.
It can be shown that $d$-separation between $A$ and $B$ by $C$ is equivalent to graph separation (in the undirected sense) in the skeleton of the moralized graph of $\G$.

	\newpage
	\section{Additional Figures}
	\label{ch13:apx:Texas}

\begin{figure}[!ht]
    \newcommand{\sfheight}{5cm}
    \centering
    \begin{subfigure}{0.34\textwidth}
        \includePlot[height=\sfheight]{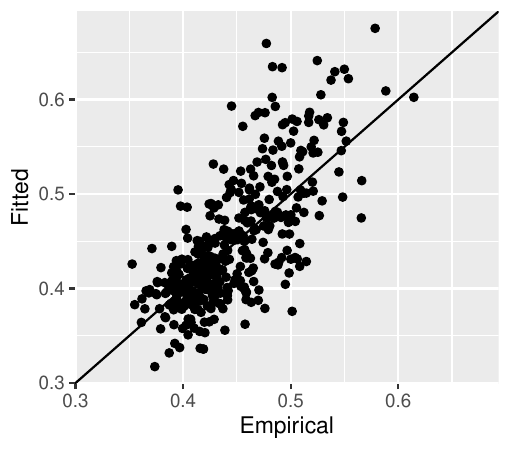}
        \subcaption{\valFlightsTexasExpertName}
        \label{ch13:fig:FlightsTexasParamsExpertKnowledge}
    \end{subfigure}
    \begin{subfigure}{0.3\textwidth}
        \includePlot[height=\sfheight]{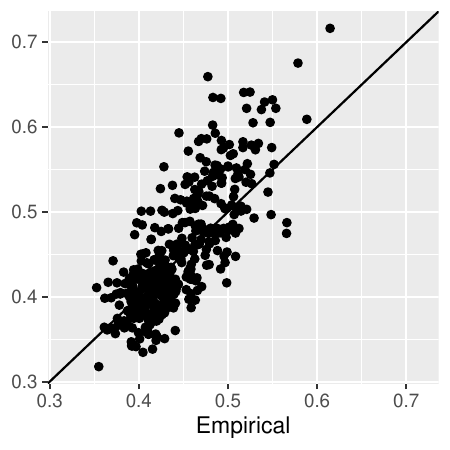}
        \subcaption{\valFullVarioMethodName}
        \label{ch13:fig:FlightsTexasParamsFullVario}
    \end{subfigure}
    \begin{subfigure}{0.3\textwidth}
        \includePlot[height=\sfheight]{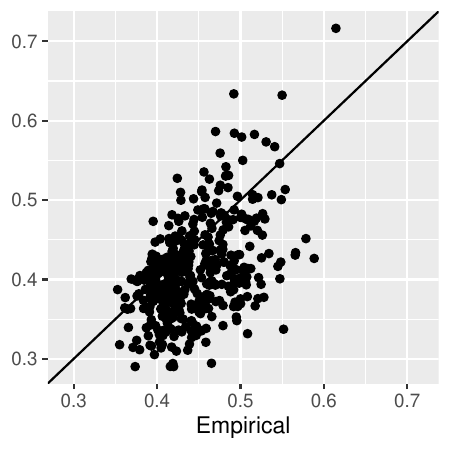}
        \subcaption{\valRandomGraphMethodName}
        \label{ch13:fig:FlightsTexasParamsRandomGraph}
    \end{subfigure}

    \begin{subfigure}{0.34\textwidth}
        \includePlot[height=\sfheight]{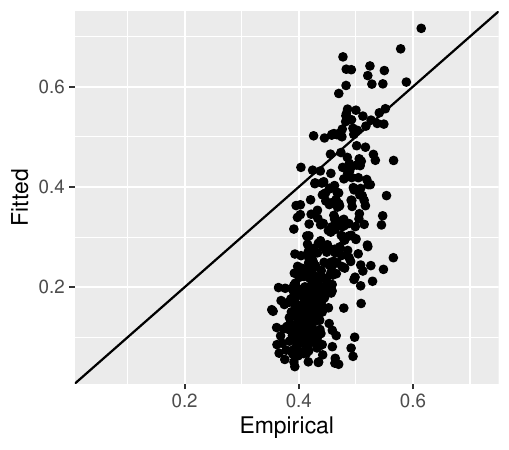}
        \subcaption{\valEmstMethodName}
        \label{ch13:fig:FlightsTexasParamsEmst}
    \end{subfigure}
    \begin{subfigure}{0.3\textwidth}
        \includePlot[height=\sfheight]{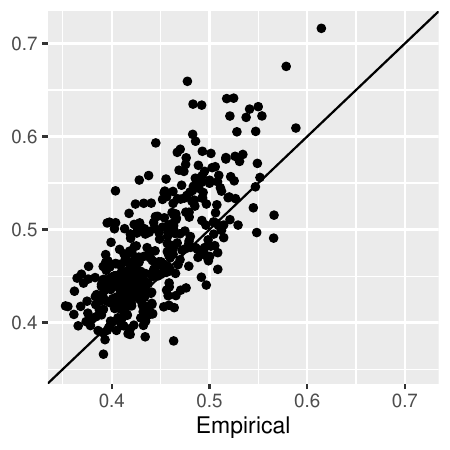}
        \subcaption{\valEglearnMethodName}
        \label{ch13:fig:FlightsTexasParamsEglearn}
    \end{subfigure}
    \begin{subfigure}{0.3\textwidth}
        \includePlot[height=\sfheight]{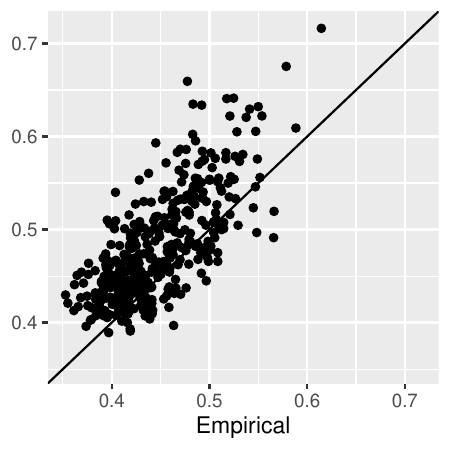}
        \subcaption{\valEmtpMethodName}
        \label{ch13:fig:FlightsTexasParamsEmtp}
    \end{subfigure}

    \begin{subfigure}{0.34\textwidth}
        \includePlot[height=\sfheight]{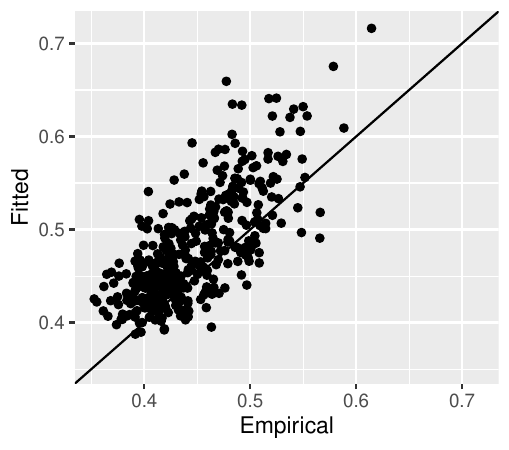}
        \subcaption{\valWazMethodName}
        \label{ch13:fig:FlightsTexasParamsWaz}
    \end{subfigure}
    \begin{subfigure}{0.3\textwidth}
        \includePlot[height=\sfheight]{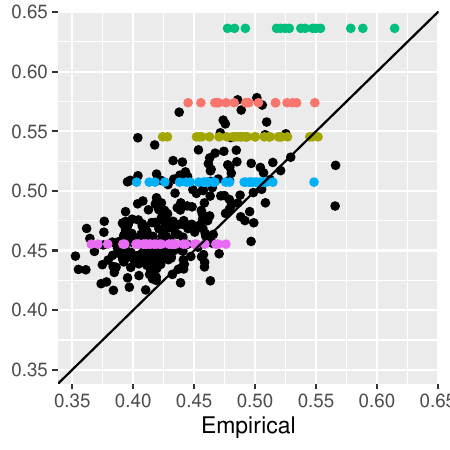}
        \subcaption{\valColorsMethodName}
        \label{ch13:fig:FlightsTexasParamsColors}
    \end{subfigure}
    \begin{subfigure}{0.3\textwidth}
        \includePlot[height=\sfheight]{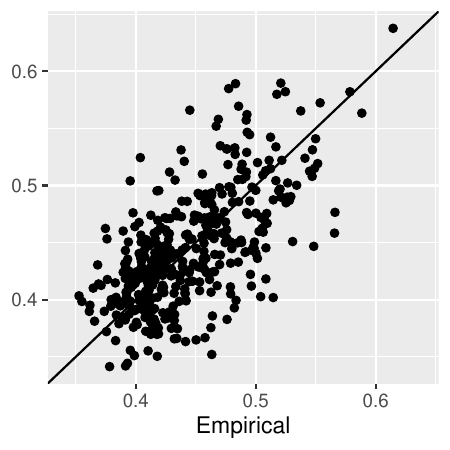}
        \subcaption{\valLeoMethodName}
        \label{ch13:fig:FlightsTexasParamsLeo}
    \end{subfigure}
    \caption{
        Extremal Correlation based on the empirical and fitted extremal variogram.
        In (\subref{ch13:fig:FlightsTexasParamsColors}),
        colored points correspond to edges of the same color.
    }
    \label{ch13:fig:FlightsTexasParams}
\end{figure}

\begin{figure}[!ht]
    \centering
    \begin{subfigure}{0.35\textwidth}
        \includePlot[width=\textwidth]{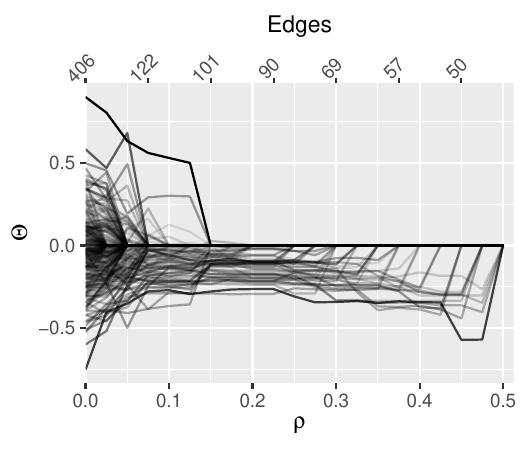}
        \subcaption{\valEglearnMethodName}
        \label{ch13:fig:FlightsTexasThetaConvergenceEglearn}
    \end{subfigure}%
    \begin{subfigure}{0.3\textwidth}
        \includePlot[width=\textwidth]{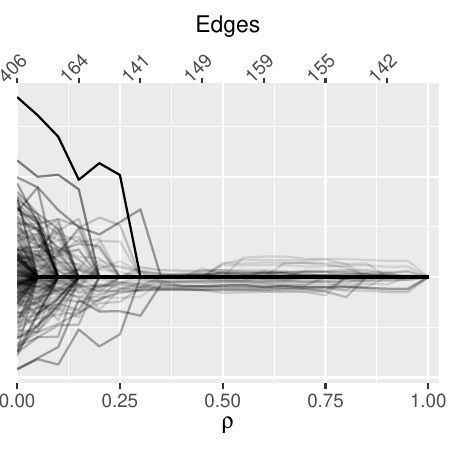}
        \subcaption{\valWazMethodName}
        \label{ch13:fig:FlightsTexasThetaConvergenceWaz}
    \end{subfigure}%
    \begin{subfigure}{0.3\textwidth}
        \includePlot[width=\textwidth]{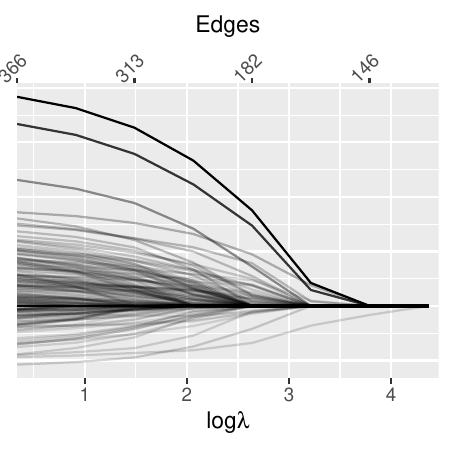}
        \subcaption{\valLeoMethodName}
        \label{ch13:fig:FlightsTexasThetaConvergenceLeo}
    \end{subfigure}%
\caption{
    Convergence of Entries in $\Theta$ to zero for penalization based methods.
    Only those entries that eventually converge to zero are shown.
}
\label{ch13:fig:FlightsTexasThetaConvergence}
\end{figure}


\begin{thebibliography}{}

\bibitem[\protect\citeauthoryear{Am{\'e}ndola, Hollering, Sullivant, and
  Tran}{Am{\'e}ndola et~al.}{2021}]{amendola2021markov}
Am{\'e}ndola, C., B.~Hollering, S.~Sullivant, and N.~Tran (2021).
\newblock Markov equivalence of max-linear {B}ayesian networks.
\newblock In {\em Uncertainty in Artificial Intelligence}, pp.\  1746--1755.
  PMLR.

\bibitem[\protect\citeauthoryear{Am\'{e}ndola, Kl\"{u}ppelberg, Lauritzen, and
  Tran}{Am\'{e}ndola et~al.}{2022}]{amendola2022conditional}
Am\'{e}ndola, C., C.~Kl\"{u}ppelberg, S.~Lauritzen, and N.~M. Tran (2022).
\newblock Conditional independence in max-linear {B}ayesian networks.
\newblock {\em Ann. Appl. Probab.\/}~{\em 32\/}(1), 1--45.

\bibitem[\protect\citeauthoryear{Asadi, Davison, and Engelke}{Asadi
  et~al.}{2015}]{asadi2015extremes}
Asadi, P., A.~C. Davison, and S.~Engelke (2015).
\newblock Extremes on river networks.
\newblock {\em Ann. Appl. Statist.\/}~{\em 9\/}(4), 2023--2050.

\bibitem[\protect\citeauthoryear{Asenova, Mazo, and Segers}{Asenova
  et~al.}{2021}]{asenova2021inference}
Asenova, S., G.~Mazo, and J.~Segers (2021).
\newblock Inference on extremal dependence in the domain of attraction of a
  structured {Hüsler--Reiss} distribution motivated by a {Markov} tree with
  latent variables.
\newblock {\em Extremes\/}~{\em 24}, 461--500.

\bibitem[\protect\citeauthoryear{Asenova and Segers}{Asenova and
  Segers}{2023}]{asenova2023extremes}
Asenova, S. and J.~Segers (2023).
\newblock Extremes of {M}arkov random fields on block graphs: max-stable limits
  and structured {H}\"{u}sler-{R}eiss distributions.
\newblock {\em Extremes\/}~{\em 26\/}(3), 433--468.

\bibitem[\protect\citeauthoryear{Banerjee, El~Ghaoui, and d'Aspremont}{Banerjee
  et~al.}{2008}]{banerjee2008}
Banerjee, O., L.~El~Ghaoui, and A.~d'Aspremont (2008).
\newblock Model selection through sparse maximum likelihood estimation for
  multivariate {G}aussian or binary data.
\newblock {\em J. Mach. Learn. Res.\/}~{\em 9}, 485--516.

\bibitem[\protect\citeauthoryear{Beirlant, Goegebeur, Segers, and
  Teugels}{Beirlant et~al.}{2006}]{beirlant2006statistics}
Beirlant, J., Y.~Goegebeur, J.~Segers, and J.~L. Teugels (2006).
\newblock {\em Statistics of extremes: theory and applications}.
\newblock John Wiley \& Sons.

\bibitem[\protect\citeauthoryear{Bortot and Coles}{Bortot and
  Coles}{2003}]{bor2003}
Bortot, P. and S.~Coles (2003).
\newblock Extremes of {M}arkov chains with tail switching potential.
\newblock {\em J. R. Stat. Soc. Ser. B Stat. Methodol.\/}~{\em 65\/}(4),
  851--867.

\bibitem[\protect\citeauthoryear{Brown and Resnick}{Brown and
  Resnick}{1977}]{bro1977}
Brown, B.~M. and S.~I. Resnick (1977).
\newblock Extreme values of independent stochastic processes.
\newblock {\em J. Appl. Probab.\/}~{\em 14}, 732--739.

\bibitem[\protect\citeauthoryear{Buck and Kl\"{u}ppelberg}{Buck and
  Kl\"{u}ppelberg}{2021}]{buc2021}
Buck, J. and C.~Kl\"{u}ppelberg (2021).
\newblock Recursive max-linear models with propagating noise.
\newblock {\em Electron. J. Stat.\/}~{\em 15\/}(2), 4770--4822.

\bibitem[\protect\citeauthoryear{Capitanio, Azzalini, and
  Stanghellini}{Capitanio et~al.}{2003}]{capitanio2003}
Capitanio, A., A.~Azzalini, and E.~Stanghellini (2003).
\newblock Graphical models for skew-normal variates.
\newblock {\em Scand. J. Statist.\/}~{\em 30\/}(1), 129--144.

\bibitem[\protect\citeauthoryear{Casey and Papastathopoulos}{Casey and
  Papastathopoulos}{2023}]{casey2023decomposable}
Casey, A. and I.~Papastathopoulos (2023).
\newblock Decomposable tail graphical models.
\newblock Available from \url{https://arxiv.org/abs/2302.05182}.

\bibitem[\protect\citeauthoryear{Chiapino, Sabourin, and Segers}{Chiapino
  et~al.}{2019}]{chiapino2019identifying}
Chiapino, M., A.~Sabourin, and J.~Segers (2019).
\newblock Identifying groups of variables with the potential of being large
  simultaneously.
\newblock {\em Extremes\/}~{\em 22}, 193--222.

\bibitem[\protect\citeauthoryear{Coles and Tawn}{Coles and
  Tawn}{1991}]{col1991}
Coles, S.~G. and J.~A. Tawn (1991).
\newblock Modelling extreme multivariate events.
\newblock {\em J. Roy. Statist. Soc. Ser. B\/}~{\em 53\/}(2), 377--392.

\bibitem[\protect\citeauthoryear{Darroch, Lauritzen, and Speed}{Darroch
  et~al.}{1980}]{darroch1980}
Darroch, J.~N., S.~L. Lauritzen, and T.~P. Speed (1980).
\newblock Markov fields and log-linear interaction models for contingency
  tables.
\newblock {\em Ann. Statist.\/}~{\em 8\/}(3), 522--539.

\bibitem[\protect\citeauthoryear{de~Fondeville and Davison}{de~Fondeville and
  Davison}{2018}]{deFondeville2018}
de~Fondeville, R. and A.~C. Davison (2018).
\newblock High-dimensional peaks-over-threshold inference.
\newblock {\em Biometrika\/}~{\em 105}, 575--592.

\bibitem[\protect\citeauthoryear{Dieker and Mikosch}{Dieker and
  Mikosch}{2015}]{die2015}
Dieker, A.~B. and T.~Mikosch (2015).
\newblock Exact simulation of {B}rown-{R}esnick random fields at a finite
  number of locations.
\newblock {\em Extremes\/}~{\em 18\/}(2), 301--314.

\bibitem[\protect\citeauthoryear{Dombry, Engelke, and Oesting}{Dombry
  et~al.}{2016}]{dombry2016}
Dombry, C., S.~Engelke, and M.~Oesting (2016).
\newblock Exact simulation of max-stable processes.
\newblock {\em Biometrika\/}~{\em 103}, 303--317.

\bibitem[\protect\citeauthoryear{Drton and Maathuis}{Drton and
  Maathuis}{2017}]{drt2017}
Drton, M. and M.~H. Maathuis (2017).
\newblock Structure learning in graphical modeling.
\newblock {\em Annu. Rev. Stat. Appl.\/}~{\em 4}, 365--393.

\bibitem[\protect\citeauthoryear{Einmahl, Kiriliouk, and Segers}{Einmahl
  et~al.}{2018}]{ein2018}
Einmahl, J. H.~J., A.~Kiriliouk, and J.~Segers (2018).
\newblock A continuous updating weighted least squares estimator of tail
  dependence in high dimensions.
\newblock {\em Extremes\/}~{\em 21\/}(2), 205--233.

\bibitem[\protect\citeauthoryear{Engelke, de~Fondeville, and Oesting}{Engelke
  et~al.}{2019}]{eng_fondeville_oesting}
Engelke, S., R.~de~Fondeville, and M.~Oesting (2019).
\newblock Extremal behaviour of aggregated data with an application to
  downscaling.
\newblock {\em Biometrika\/}~{\em 106\/}(1), 127--144.

\bibitem[\protect\citeauthoryear{Engelke and Hitz}{Engelke and
  Hitz}{2020}]{engelkeHitz2020}
Engelke, S. and A.~S. Hitz (2020).
\newblock Graphical models for extremes (with discussion).
\newblock {\em J. R. Stat. Soc. Ser. B Stat. Methodol\/}~{\em 82\/}(4),
  871--932.

\bibitem[\protect\citeauthoryear{Engelke, Hitz, Gnecco, and Hentschel}{Engelke
  et~al.}{2022}]{graphicalExtremes2022}
Engelke, S., A.~S. Hitz, N.~Gnecco, and M.~Hentschel (2022).
\newblock {\em graphicalExtremes: Statistical Methodology for Graphical Extreme
  Value Models}.
\newblock Available from
  \url{https://github.com/sebastian-engelke/graphicalExtremes}, R package
  version 0.1.0.9000.

\bibitem[\protect\citeauthoryear{Engelke and Ivanovs}{Engelke and
  Ivanovs}{2021}]{engelke2021a}
Engelke, S. and J.~Ivanovs (2021).
\newblock Sparse structures for multivariate extremes.
\newblock {\em Annu. Rev. Stat. Appl.\/}~{\em 8}, 241--270.

\bibitem[\protect\citeauthoryear{Engelke, Ivanovs, and Strokorb}{Engelke
  et~al.}{2022}]{engelkeIvanovsStrokorb2022}
Engelke, S., J.~Ivanovs, and K.~Strokorb (2022).
\newblock Graphical models for infinite measures with applications to extremes
  and {L}\'evy processes.
\newblock Available from \url{https://arxiv.org/abs/2211.15769}.

\bibitem[\protect\citeauthoryear{Engelke, Lalancette, and Volgushev}{Engelke
  et~al.}{2022}]{engelke2022a}
Engelke, S., M.~Lalancette, and S.~Volgushev (2022).
\newblock Learning extremal graphical structures in high dimensions.
\newblock Available from \url{https://arxiv.org/abs/2111.00840}.

\bibitem[\protect\citeauthoryear{Engelke, Malinowski, Kabluchko, and
  Schlather}{Engelke et~al.}{2015}]{engelke2014}
Engelke, S., A.~Malinowski, Z.~Kabluchko, and M.~Schlather (2015, may).
\newblock Estimation of {H}üsler-{R}eiss distributions and {B}rown-{R}esnick
  processes.
\newblock {\em J. R. Stat. Soc. Ser. B Stat. Methodol\/}~{\em 77\/}(1),
  239--265.

\bibitem[\protect\citeauthoryear{Engelke, Malinowski, Oesting, and
  Schlather}{Engelke et~al.}{2014}]{eng2014}
Engelke, S., A.~Malinowski, M.~Oesting, and M.~Schlather (2014).
\newblock Statistical inference for max-stable processes by conditioning on
  extreme events.
\newblock {\em Adv. in Appl. Probab.\/}~{\em 46\/}(2), 478--495.

\bibitem[\protect\citeauthoryear{Engelke and Volgushev}{Engelke and
  Volgushev}{2022}]{engelkeVolgushev2022}
Engelke, S. and S.~Volgushev (2022).
\newblock Structure learning for extremal tree models.
\newblock {\em J. R. Stat. Soc. Ser. B. Stat. Methodol.\/}~{\em 84\/}(5),
  2055--2087.

\bibitem[\protect\citeauthoryear{Fiebig, Strokorb, and Schlather}{Fiebig
  et~al.}{2017}]{fie2017}
Fiebig, U.-R., K.~Strokorb, and M.~Schlather (2017).
\newblock The realization problem for tail correlation functions.
\newblock {\em Extremes\/}~{\em 20\/}(1), 121--168.

\bibitem[\protect\citeauthoryear{Friedman, Hastie, and Tibshirani}{Friedman
  et~al.}{2008}]{friedmanEtAl2007}
Friedman, J., T.~Hastie, and R.~Tibshirani (2008, 12).
\newblock {Sparse inverse covariance estimation with the graphical lasso}.
\newblock {\em Biostatistics\/}~{\em 9\/}(3), 432--441.

\bibitem[\protect\citeauthoryear{Geiger, Verma, and Pearl}{Geiger
  et~al.}{1990}]{GVP1990}
Geiger, D., T.~Verma, and J.~Pearl (1990).
\newblock Identifying independence in {B}ayesian networks.
\newblock {\em Networks\/}~{\em 20\/}(5), 507--534.

\bibitem[\protect\citeauthoryear{Gissibl and Kl\"{u}ppelberg}{Gissibl and
  Kl\"{u}ppelberg}{2018}]{gissibl2018}
Gissibl, N. and C.~Kl\"{u}ppelberg (2018).
\newblock Max-linear models on directed acyclic graphs.
\newblock {\em Bernoulli\/}~{\em 24\/}(4A), 2693--2720.

\bibitem[\protect\citeauthoryear{Gissibl, Kl\"{u}ppelberg, and
  Lauritzen}{Gissibl et~al.}{2021}]{gissibl2021identifiability}
Gissibl, N., C.~Kl\"{u}ppelberg, and S.~Lauritzen (2021).
\newblock Identifiability and estimation of recursive max-linear models.
\newblock {\em Scand. J. Stat.\/}~{\em 48\/}(1), 188--211.

\bibitem[\protect\citeauthoryear{Gnecco, Meinshausen, Peters, and
  Engelke}{Gnecco et~al.}{2021}]{gneccoEtAl2021}
Gnecco, N., N.~Meinshausen, J.~Peters, and S.~Engelke (2021).
\newblock {Causal discovery in heavy-tailed models}.
\newblock {\em Ann. Statist.\/}~{\em 49\/}(3), 1755 -- 1778.

\bibitem[\protect\citeauthoryear{Gudendorf and Segers}{Gudendorf and
  Segers}{2010}]{gudendorf2010extreme}
Gudendorf, G. and J.~Segers (2010).
\newblock Extreme-value copulas.
\newblock In {\em Copula Theory and Its Applications: Proceedings of the
  Workshop Held in Warsaw, 25-26 September 2009}, pp.\  127--145. Springer.

\bibitem[\protect\citeauthoryear{Gumbel}{Gumbel}{1960}]{gumbel1960}
Gumbel, E.~J. (1960).
\newblock Distributions des valeurs extr\^{e}mes en plusieurs dimensions.
\newblock {\em Publ. Inst. Statist. Univ. Paris\/}~{\em 9}, 171--173.

\bibitem[\protect\citeauthoryear{Heffernan and Tawn}{Heffernan and
  Tawn}{2004}]{heffernan2004}
Heffernan, J.~E. and J.~A. Tawn (2004).
\newblock A conditional approach for multivariate extreme values.
\newblock {\em J. R. Stat. Soc. Ser. B Stat. Methodol\/}~{\em 66\/}(3),
  497--546.

\bibitem[\protect\citeauthoryear{Hentschel, Engelke, and Segers}{Hentschel
  et~al.}{2022}]{hentschelEngelkeSegers2022}
Hentschel, M., S.~Engelke, and J.~Segers (2022).
\newblock Statistical inference for h\"usler-reiss graphical models through
  matrix completions.

\bibitem[\protect\citeauthoryear{H{\"u}sler and Reiss}{H{\"u}sler and
  Reiss}{1989}]{hueslerReiss1989}
H{\"u}sler, J. and R.-D. Reiss (1989, February).
\newblock {Maxima of normal random vectors: Between independence and complete
  dependence}.
\newblock {\em Statist. Prob. Letters\/}~{\em 7\/}(4), 283--286.

\bibitem[\protect\citeauthoryear{Hyv\"{a}rinen}{Hyv\"{a}rinen}{2005}]{Hyvaerinen2005}
Hyv\"{a}rinen, A. (2005).
\newblock Estimation of non-normalized statistical models by score matching.
\newblock {\em J. Mach. Learn. Res.\/}~{\em 6}, 695--709.

\bibitem[\protect\citeauthoryear{Inouye, Ravikumar, and Dhillon}{Inouye
  et~al.}{2016}]{inouye2016square}
Inouye, D., P.~Ravikumar, and I.~Dhillon (2016).
\newblock Square root graphical models: Multivariate generalizations of
  univariate exponential families that permit positive dependencies.
\newblock In {\em International conference on machine learning}, pp.\
  2445--2453. PMLR.

\bibitem[\protect\citeauthoryear{Kabluchko, Schlather, and de~Haan}{Kabluchko
  et~al.}{2009}]{kabluchkoSchlatherDeHaan2009}
Kabluchko, Z., M.~Schlather, and L.~de~Haan (2009).
\newblock {Stationary max-stable fields associated to negative definite
  functions}.
\newblock {\em Ann. Probab.\/}~{\em 37\/}(5), 2042 -- 2065.

\bibitem[\protect\citeauthoryear{Kiriliouk, Lee, and Segers}{Kiriliouk
  et~al.}{2023}]{kiriliouk2023xvine}
Kiriliouk, A., J.~Lee, and J.~Segers (2023).
\newblock X-vine models for multivariate extremes.

\bibitem[\protect\citeauthoryear{Klein, Orellana, Brincat, Miller, and
  Kass}{Klein et~al.}{2020}]{KOBMK20}
Klein, N., J.~Orellana, S.~L. Brincat, E.~K. Miller, and R.~E. Kass (2020).
\newblock Torus graphs for multivariate phase coupling analysis.
\newblock {\em Ann. Appl. Stat.\/}~{\em 14\/}(2), 635--660.

\bibitem[\protect\citeauthoryear{Kl\"{u}ppelberg and Krali}{Kl\"{u}ppelberg and
  Krali}{2021}]{klu2021scaling}
Kl\"{u}ppelberg, C. and M.~Krali (2021).
\newblock Estimating an extreme {B}ayesian network via scalings.
\newblock {\em J. Multivariate Anal.\/}~{\em 181}, Paper No. 104672, 23.

\bibitem[\protect\citeauthoryear{Krali, Davison, and Kl{\"u}ppelberg}{Krali
  et~al.}{2023}]{krali2023heavy}
Krali, M., A.~C. Davison, and C.~Kl{\"u}ppelberg (2023).
\newblock Heavy-tailed max-linear structural equation models in networks with
  hidden nodes.
\newblock Available from \url{https://arxiv.org/abs/2306.15356}.

\bibitem[\protect\citeauthoryear{Kruskal}{Kruskal}{1956}]{kruskal1956shortest}
Kruskal, Jr., J.~B. (1956).
\newblock On the shortest spanning subtree of a graph and the traveling
  salesman problem.
\newblock {\em Proc. Amer. Math. Soc.\/}~{\em 7}, 48--50.

\bibitem[\protect\citeauthoryear{Kulik and Soulier}{Kulik and
  Soulier}{2015}]{kulik2015}
Kulik, R. and P.~Soulier (2015).
\newblock Heavy tailed time series with extremal independence.
\newblock {\em Extremes\/}~{\em 18\/}(2), 273--299.

\bibitem[\protect\citeauthoryear{Lalancette}{Lalancette}{2023}]{lalancette2023pairwise}
Lalancette, M. (2023).
\newblock On pairwise interaction multivariate {P}areto models.
\newblock {\em Stat\/}~{\em 12}, Paper No. e613, 10.

\bibitem[\protect\citeauthoryear{Lauritzen and Sadeghi}{Lauritzen and
  Sadeghi}{2018}]{LS2018}
Lauritzen, S. and K.~Sadeghi (2018).
\newblock Unifying {M}arkov properties for graphical models.
\newblock {\em Ann. Statist.\/}~{\em 46\/}(5), 2251--2278.

\bibitem[\protect\citeauthoryear{Lauritzen and Zwiernik}{Lauritzen and
  Zwiernik}{2022}]{LZ2022}
Lauritzen, S. and P.~Zwiernik (2022).
\newblock Locally associated graphical models and mixed convex exponential
  families.
\newblock {\em Ann. Statist.\/}~{\em 50\/}(5), 3009--3038.

\bibitem[\protect\citeauthoryear{Lauritzen}{Lauritzen}{1996}]{lauritzen1996}
Lauritzen, S.~L. (1996).
\newblock {\em Graphical models}, Volume~17 of {\em Oxford statistical science
  series}.
\newblock Oxford: Clarendon Press.

\bibitem[\protect\citeauthoryear{Lederer and Oesting}{Lederer and
  Oesting}{2023}]{lederer2023extremes}
Lederer, J. and M.~Oesting (2023).
\newblock Extremes in high dimensions: Methods and scalable algorithms.
\newblock Available from \url{https://arxiv.org/abs/2303.04258}.

\bibitem[\protect\citeauthoryear{Lee and Joe}{Lee and Joe}{2018}]{lee2018}
Lee, D. and H.~Joe (2018).
\newblock Multivariate extreme value copulas with factor and tree dependence
  structures.
\newblock {\em Extremes\/}~{\em 21\/}(1), 147--176.

\bibitem[\protect\citeauthoryear{Lin, Drton, and Shojaie}{Lin
  et~al.}{2016}]{LDS16}
Lin, L., M.~Drton, and A.~Shojaie (2016).
\newblock Estimation of high-dimensional graphical models using regularized
  score matching.
\newblock {\em Electron. J. Stat.\/}~{\em 10\/}(1), 806--854.

\bibitem[\protect\citeauthoryear{Liu, Lafferty, and Wasserman}{Liu
  et~al.}{2009}]{LLW09}
Liu, H., J.~Lafferty, and L.~Wasserman (2009).
\newblock The nonparanormal: semiparametric estimation of high dimensional
  undirected graphs.
\newblock {\em J. Mach. Learn. Res.\/}~{\em 10}, 2295--2328.

\bibitem[\protect\citeauthoryear{Liu, Blanchet, Dieker, and Mikosch}{Liu
  et~al.}{2019}]{zhi2019}
Liu, Z., J.~H. Blanchet, A.~B. Dieker, and T.~Mikosch (2019).
\newblock On logarithmically optimal exact simulation of max-stable and related
  random fields on a compact set.
\newblock {\em Bernoulli\/}~{\em 25\/}(4A), 2949--2981.

\bibitem[\protect\citeauthoryear{Maathuis, Drton, Lauritzen, and
  Wainwright}{Maathuis et~al.}{2019}]{maathuis2019handbook}
Maathuis, M., M.~Drton, S.~Lauritzen, and M.~Wainwright (Eds.) (2019).
\newblock {\em Handbook of graphical models}.
\newblock Chapman \& Hall/CRC Handbooks of Modern Statistical Methods. CRC
  Press, Boca Raton, FL.

\bibitem[\protect\citeauthoryear{Meinshausen and B{\"u}hlmann}{Meinshausen and
  B{\"u}hlmann}{2006}]{meinshausen2006}
Meinshausen, N. and P.~B{\"u}hlmann (2006).
\newblock High-dimensional graphs and variable selection with the lasso.
\newblock {\em Ann. Stat.\/}~{\em 34\/}(3), 1436--1462.

\bibitem[\protect\citeauthoryear{Meyer and Wintenberger}{Meyer and
  Wintenberger}{2023}]{meyer2023multivariate}
Meyer, N. and O.~Wintenberger (2023).
\newblock Multivariate sparse clustering for extremes.
\newblock {\em J. Am. Stat. Assoc.\/}~(forthcoming), 1--23.

\bibitem[\protect\citeauthoryear{Opitz}{Opitz}{2013}]{opi2013}
Opitz, T. (2013).
\newblock Extremal {$t$} processes: Elliptical domain of attraction and a
  spectral representation.
\newblock {\em J. Multivariate Anal.\/}~{\em 122}, 409--413.

\bibitem[\protect\citeauthoryear{Papastathopoulos, Casey, and
  Tawn}{Papastathopoulos et~al.}{2023}]{papastathopoulos2023hidden}
Papastathopoulos, I., A.~Casey, and J.~A. Tawn (2023).
\newblock Hidden tail chains and recurrence equations for dependence parameters
  associated with extremes of higher-order {M}arkov chains.
\newblock Available from \url{https://arxiv.org/abs/1903.04059}.

\bibitem[\protect\citeauthoryear{Papastathopoulos and
  Strokorb}{Papastathopoulos and Strokorb}{2016}]{papastathopoulosStrokorb2016}
Papastathopoulos, I. and K.~Strokorb (2016).
\newblock Conditional independence among max-stable laws.
\newblock {\em Statistics \& Probability Letters\/}~{\em 108}, 9--15.

\bibitem[\protect\citeauthoryear{Papastathopoulos, Strokorb, Tawn, and
  Butler}{Papastathopoulos et~al.}{2017}]{papastathopoulos2017}
Papastathopoulos, I., K.~Strokorb, J.~A. Tawn, and A.~Butler (2017).
\newblock Extreme events of {Markov} chains.
\newblock {\em Adv. in Appl. Probab.\/}~{\em 49\/}(1), 134--161.

\bibitem[\protect\citeauthoryear{Pearl}{Pearl}{1988}]{pearl1988}
Pearl, J. (1988).
\newblock {\em Probabilistic reasoning in intelligent systems: networks of
  plausible inference}.
\newblock Morgan Kaufmann.

\bibitem[\protect\citeauthoryear{Pearl}{Pearl}{2009}]{Pearl2000}
Pearl, J. (2009).
\newblock {\em Causality\/} (Second ed.).
\newblock Cambridge University Press, Cambridge.
\newblock Models, reasoning, and inference.

\bibitem[\protect\citeauthoryear{Perfekt}{Perfekt}{1994}]{per1994}
Perfekt, R. (1994).
\newblock Extremal behaviour of stationary {M}arkov chains with applications.
\newblock {\em Ann. Appl. Probab.\/}~{\em 4\/}(2), 529--548.

\bibitem[\protect\citeauthoryear{Peters, Janzing, and Sch\"{o}lkopf}{Peters
  et~al.}{2017}]{PJS2017}
Peters, J., D.~Janzing, and B.~Sch\"{o}lkopf (2017).
\newblock {\em Elements of causal inference}.
\newblock Adaptive Computation and Machine Learning. MIT Press, Cambridge, MA.
\newblock Foundations and learning algorithms.

\bibitem[\protect\citeauthoryear{Prim}{Prim}{1957}]{pri1957}
Prim, R.~C. (1957).
\newblock Shortest connection networks and some generalizations.
\newblock {\em Bell System Technical Journal\/}~{\em 36}, 1389--1401.

\bibitem[\protect\citeauthoryear{Ravikumar, Wainwright, and Lafferty}{Ravikumar
  et~al.}{2010}]{pra2010}
Ravikumar, P., M.~J. Wainwright, and J.~D. Lafferty (2010).
\newblock High-dimensional {I}sing model selection using {$\ell_1$}-regularized
  logistic regression.
\newblock {\em Ann. Statist.\/}~{\em 38\/}(3), 1287--1319.

\bibitem[\protect\citeauthoryear{Resnick}{Resnick}{2008}]{res2008}
Resnick, S.~I. (2008).
\newblock {\em Extreme Values, Regular Variation and Point Processes}.
\newblock New York: Springer.

\bibitem[\protect\citeauthoryear{R\"{o}ttger, Engelke, and
  Zwiernik}{R\"{o}ttger et~al.}{2023}]{REZ2023}
R\"{o}ttger, F., S.~Engelke, and P.~Zwiernik (2023).
\newblock Total positivity in multivariate extremes.
\newblock {\em Ann. Statist.\/}~{\em 51\/}(3), 962--1004.

\bibitem[\protect\citeauthoryear{Röttger, Coons, and Grosdos}{Röttger
  et~al.}{2023}]{roettger2023parametric}
Röttger, F., J.~I. Coons, and A.~Grosdos (2023).
\newblock Parametric and nonparametric symmetries in graphical models for
  extremes.
\newblock Available from \url{https://arxiv.org/abs/2306.00703}.

\bibitem[\protect\citeauthoryear{Röttger and Schmitz}{Röttger and
  Schmitz}{2023}]{roettgerSchmitz2023}
Röttger, F. and Q.~Schmitz (2023).
\newblock On the local metric property in multivariate extremes.
\newblock Available from \url{https://arxiv.org/abs/2212.10350}.

\bibitem[\protect\citeauthoryear{Schlather}{Schlather}{2002}]{schlater2002}
Schlather, M. (2002).
\newblock Models for stationary max-stable random fields.
\newblock {\em Extremes\/}~{\em 5\/}(1), 33--44.

\bibitem[\protect\citeauthoryear{Segers}{Segers}{2007}]{segers2007multivariate}
Segers, J. (2007).
\newblock Multivariate regular variation of heavy-tailed {M}arkov chains.
\newblock Available from \url{https://arxiv.org/abs/math/0701411}.

\bibitem[\protect\citeauthoryear{Segers}{Segers}{2020}]{segers2020}
Segers, J. (2020).
\newblock One- versus multi-component regular variation and extremes of
  {Markov} trees.
\newblock {\em Adv. in Appl. Probab.\/}~{\em 52}, 855--878.

\bibitem[\protect\citeauthoryear{Segers and Asenova}{Segers and
  Asenova}{2022}]{segers2022max}
Segers, J. and S.~Asenova (2022).
\newblock Max-linear graphical models with heavy-tailed factors on trees of
  transitive tournaments.
\newblock Available from \url{https://arxiv.org/abs/2209.14938}.

\bibitem[\protect\citeauthoryear{Simpson, Wadsworth, and Tawn}{Simpson
  et~al.}{2020}]{simpson2020determining}
Simpson, E.~S., J.~L. Wadsworth, and J.~A. Tawn (2020, 05).
\newblock {Determining the dependence structure of multivariate extremes}.
\newblock {\em Biometrika\/}~{\em 107\/}(3), 513--532.

\bibitem[\protect\citeauthoryear{Smith}{Smith}{1992}]{smi1992}
Smith, R.~L. (1992).
\newblock The extremal index for a {M}arkov chain.
\newblock {\em J. Appl. Probab.\/}~{\em 29\/}(1), 37--45.

\bibitem[\protect\citeauthoryear{Smith, Tawn, and Coles}{Smith
  et~al.}{1997}]{smi1997}
Smith, R.~L., J.~A. Tawn, and S.~G. Coles (1997).
\newblock Markov chain models for threshold exceedances.
\newblock {\em Biometrika\/}~{\em 84\/}(2), 249--268.

\bibitem[\protect\citeauthoryear{Strokorb}{Strokorb}{2020}]{strokorb2020extremal}
Strokorb, K. (2020).
\newblock Extremal independence old and new.
\newblock Available from \url{https://arxiv.org/abs/2002.07808}.

\bibitem[\protect\citeauthoryear{Thibaud, Mutzner, and Davison}{Thibaud
  et~al.}{2013}]{thibaud_davison}
Thibaud, E., R.~Mutzner, and A.~C. Davison (2013).
\newblock Threshold modeling of extreme spatial rainfall.
\newblock {\em Water Resources Research\/}~{\em 49\/}(8), 4633--4644.

\bibitem[\protect\citeauthoryear{Tran, Buck, and Kl{\"u}ppelberg}{Tran
  et~al.}{2021}]{tran2021estimating}
Tran, N.~M., J.~Buck, and C.~Kl{\"u}ppelberg (2021).
\newblock Estimating a directed tree for extremes.
\newblock Available from \url{https://arxiv.org/abs/2102.06197}.

\bibitem[\protect\citeauthoryear{Vettori, Huser, Segers, and Genton}{Vettori
  et~al.}{2020}]{vet2020}
Vettori, S., R.~Huser, J.~Segers, and M.~G. Genton (2020).
\newblock Bayesian model averaging over tree-based dependence structures for
  multivariate extremes.
\newblock {\em J. Comput. Graph. Statist.\/}~{\em 29\/}(1), 174--190.

\bibitem[\protect\citeauthoryear{Wainwright and Jordan}{Wainwright and
  Jordan}{2008}]{wainwright2008}
Wainwright, M. and M.~Jordan (2008).
\newblock Graphical models, exponential families, and variational inference.
\newblock {\em Foundations and Trends in Machine Learning\/}~{\em 1\/}(1--2),
  1--305.

\bibitem[\protect\citeauthoryear{Wan and Zhou}{Wan and
  Zhou}{2023}]{wan2023graphical}
Wan, P. and C.~Zhou (2023).
\newblock Graphical lasso for extremes.
\newblock Available from \url{https://arxiv.org/abs/2307.15004}.

\bibitem[\protect\citeauthoryear{Yang, Ravikumar, Allen, and Liu}{Yang
  et~al.}{2015}]{YRAL15}
Yang, E., P.~Ravikumar, G.~I. Allen, and Z.~Liu (2015).
\newblock Graphical models via univariate exponential family distributions.
\newblock {\em J. Mach. Learn. Res.\/}~{\em 16}, 3813--3847.

\bibitem[\protect\citeauthoryear{Yang, Ravikumar, Allen, and Liu}{Yang
  et~al.}{2013}]{yang2013poisson}
Yang, E., P.~K. Ravikumar, G.~I. Allen, and Z.~Liu (2013).
\newblock On poisson graphical models.
\newblock {\em Advances in neural information processing systems\/}~{\em 26}.

\bibitem[\protect\citeauthoryear{Ying, Cardoso, and Palomar}{Ying
  et~al.}{2021}]{ying2021minimax}
Ying, J., J.~M. Cardoso, and D.~Palomar (2021).
\newblock Minimax estimation of {L}aplacian constrained precision matrices.
\newblock In {\em Int. Conf. Artif. Intell. Stat.}, pp.\  3736--3744. PMLR.

\bibitem[\protect\citeauthoryear{Ying, Cardoso, and Palomar}{Ying
  et~al.}{2020}]{ying2020does}
Ying, J., J.~V. d.~M. Cardoso, and D.~P. Palomar (2020).
\newblock Does the $\ell_1$-norm learn a sparse graph under {L}aplacian
  constrained graphical models?
\newblock Available from \url{https://arxiv.org/abs/2006.14925}.

\bibitem[\protect\citeauthoryear{Yuan and Lin}{Yuan and
  Lin}{2007}]{yuan2007model}
Yuan, M. and Y.~Lin (2007).
\newblock Model selection and estimation in the {G}aussian graphical model.
\newblock {\em Biometrika\/}~{\em 94\/}(1), 19--35.

\end{thebibliography}

\end{document}